\documentstyle[12pt,aaspp4]{article}

\tighten

\journalid{337}{15 January 1989}
\articleid{11}{14}

\slugcomment{Accepted for publication in ApJS}

\lefthead{Tucker et al.}
\righthead{Loose Groups of Galaxies in the LCRS}

\begin{document}

\title{Loose Groups of Galaxies in the Las Campanas Redshift Survey}

\author{Douglas L.\ Tucker}
\affil{Fermi National Accelerator Laboratory,
       MS~127, 
       P.O.~Box~500, 
       Batavia, IL 60510, 
       USA; 
       dtucker@fnal.gov}

\author{Augustus Oemler, Jr.\altaffilmark{1}, 
        Yasuhiro Hashimoto\altaffilmark{1,2}, and
        Stephen A.\ Shectman}
\affil{Carnegie Observatories, 
       813 Santa Barbara Street, 
       Pasadena, CA 91101, 
       USA; 
       oemler@ociw.edu, hashimoto@vorpal.ociw.edu, shec@ociw.edu}
\altaffiltext{1}{Also:  Dept.\ of Astronomy, 
                        Yale University, 
                        New Haven, CT 06520-8101, USA.}
\altaffiltext{2}{Present Address:  Astrophysikalisches Institut Potsdam,
                                   An der Sternwarte 16, 
                                   D-14482 Potsdam, Germany.}

\author{Robert P.\ Kirshner}
\affil{Harvard-Smithsonian Center for Astrophysics, 
       60 Garden Street,
       Cambridge, MA 02138, 
       USA;
       kirshner@cfa.harvard.edu}

\author{Huan Lin\altaffilmark{3}}
\affil{Steward Observatory,
       University of Arizona, 
       933 N. Cherry Ave., 
       Tucson, AZ 85721,
       USA;
       hlin@as.arizona.edu}
\altaffiltext{3}{Hubble Fellow.}

\author{Stephen D.\ Landy}
\affil{Dept.\ of Physics,
       College of William \& Mary,
       Williamsburg, VA 23187, 
       USA; 
       landy@physics.wm.edu}

\author{Paul L.\ Schechter}
\affil{Dept.\ of Physics,
       Massachusetts Institute of Technology, 
       Cambridge, MA 02139, 
       USA; 
       schech@achernar.mit.edu}

\and

\author{Sahar S. Allam\altaffilmark{4}}
\affil{National Research Institute for Astronomy \& Geophysics,
       Helwan Observatory, 
       Cairo, 
       Egypt; 
       shr@frcu.eun.eg}
\altaffiltext{4}{Visiting Scientist, Fermi National Accelerator Laboratory.}

\begin{abstract}
A ``friends-of-friends'' percolation algorithm has been used to
extract a catalogue of $\delta n / n = 80$ density enhancements
(groups) from the six slices of the Las Campanas Redshift Survey
(LCRS).  The full catalogue contains 1495 groups and includes 35\% of
the LCRS galaxy sample.  A clean sample of 394 groups has been
derived by culling groups from the full sample which either are too
close to a slice edge, have a crossing time greater than a Hubble
time, have a corrected velocity dispersion of zero, or contain
a 55-arcsec ``orphan'' (a galaxy with a mock redshift which was excluded
from the original LCRS redshift catalogue due to its proximity to
another galaxy --- i.e., within 55~arcsec).  Median properties derived
from the clean sample include: line-of-sight velocity dispersion
$\sigma_{\rm los} = 164$~km~s$^{-1}$, crossing time $t_{\rm cr} =
0.10~H_0^{-1}$, harmonic radius $R_{\rm h} = 0.58~h^{-1}$~Mpc,
pairwise separation $R_{\rm p} = 0.64~h^{-1}$~Mpc, virial mass $M_{\rm
vir} = 1.90~\times~10^{13}~h^{-1}~M_{\sun}$, total group $R$-band
luminosity $L_{\rm tot} = 1.30~\times~10^{11}~h^{-2}~L_{\sun}$, and
$R$-band mass-to-light ratio $M/L = 171~h~M_{\sun}/L_{\sun}$; the
median number of observed members in a group is 3.
\end{abstract}


\section{Introduction to Group Catalogues}

Loose groups of galaxies are important but little-understood entities.
They are intermediate in scale between galaxies and rich clusters, and
thus their dynamics are important in the study of the distribution of
dark matter on scales greater than haloes of galaxies but smaller than
the typical sizes of large clusters [see, for example, the review by
Oemler (1988)].  Their environment is also intermediate between that
of isolated galaxies and that of the cores of rich clusters, and
therefore the study of groups may provide clues to the processes that
create the observed dependency of galaxy morphology on environment
(Postman \& Geller 1984; Oemler 1992; Allington-Smith et al.\ 1993;
Whitmore, Gilmore, \& Jones 1993; Zabludoff et al. 1996; Hashimoto et
al.\ 1998).  Only in the past 15 years, however, with the advent of
extensive galaxy redshift surveys, have suitably uncontaminated,
objective group catalogues been available for study.

Galaxies and rich clusters of galaxies are generally easy to identify.
They are high-contrast objects compared with their immediate
surroundings.  Unfortunately, loose groups of galaxies, which are
neither particularly dense nor exceptionally populous, are much more
difficult to distinguish from their surroundings.  Early group
catalogues were based upon the identification of galaxy concentrations
on the sky, first primarily by visual inspection of photographic
plates (e.g., Holmberg 1969, de~Vaucouleurs 1975), and, later, via
objective group-finding algorithms (e.g., Turner \& Gott 1976).  Since
these group catalogues relied especially on just the two dimensions of
spatial information available on the plane of the sky, they were
greatly subject to contamination from projection effects.  Projection
effects are largely mitigated (although never fully eliminated) in
group catalogues derived from galaxy redshift surveys.  In the
early-1980's, Huchra \& Geller (1982; HG82), pioneers in the
extraction of groups from redshift surveys by means of objective,
``friends-of-friends'' percolation algorithms, compiled a group
catalogue from a shallow ($m_{B(0)}^{\rm lim} = 13.2$) whole-sky
redshift catalogue containing 1312 galaxies.  They later derived a
group catalogue from the original ($m_{B(0)}^{\rm lim} = 14.5$) CfA
Survey (hereafter, CfA1) (Geller \& Huchra 1983; GH83).  The CfA1 has
in fact proved to be a popular testing ground for group-finding
algorithms; additional group catalogues drawn from the CfA1 include
those by Nolthenius \& White (1987; NW87), Nolthenius (1993; N93), and
Moore, Frenk, \& White (1993; MFW93). Groups have also been identified
in a $12\arcdeg$ slice from the CfA extension to $m_{B(0)}^{\rm lim} =
15.5$ [henceforth, CfA2; Ramella, Geller, \& Huchra 1989 (RGH89)] and
in the diameter-limited Southern Sky Redshift Survey [SSRS; Maia, da
Costa, \& Latham 1989 (MdCL89)].  More recently, group catalogues have
been extracted from the full northern CfA2 by Ramella, Pisani, \&
Geller (1997; RPG97), from the Pisces-Perseus redshift survey (PPS) by
Trasarti-Battistoni (1998; TB98), and from the ESO Slice Project (ESP)
galaxy redshift survey by Ramella et al.\ (1999; RZZ99).

In this paper, we will present a group catalogue based upon the Las
Campanas Redshift Survey (LCRS; Shectman et al.\ 1996).  Due to the
large volume this survey samples, the LCRS includes numerous ``Great
Wall''-like structures within its borders and is therefore one of the
first redshift surveys which can claim to enclose a reasonably fair
sample of the nearby Universe.  With the exception of the ESP, the
redshift surveys from which the aforementioned group catalogues have
been derived have all tended to be dominated by a very few large
structures.  Therefore, a group catalogue based upon the LCRS should
be found to contain groups in a wider range of environments than the
groups identified from these shallower surveys.  A census of group
properties based upon LCRS groups would thus be more complete, and
therefore more useful for studies of both galaxy dynamics and
environmental dependences.  In fact, this characteristic is so
important that earlier variations of the present catalogue have
already been used in studies of the environmental influence on galaxy
morphology (Hashimoto \& Oemler 1998), on the presence of ``E+A''
galaxies (Zabludoff et al.\ 1996), and on the general rate of star
formation within galaxies (Hashimoto et al.\ 1998, Allam et al.\
1999).

We divide the remainder of this paper as follows: we describe the LCRS
galaxy sample in \S~2, discuss the modified ``friends-of-friends''
algorithm used to extract the LCRS group catalogue in \S~3, present
the catalogue itself in \S~4, and compare it with various group
catalogues and with Abell clusters [Abell 1958; Abell, Corwin, \&
Olowin 1989 (ACO)] in \S~5 and \S~6; in \S~7, we summarize and
conclude.

\section{The Data}

The LCRS is an optically selected galaxy redshift survey which
extends to a redshift of 0.2 and which is composed of a total of 6
alternating $1.5^{\arcdeg} \times 80^{\arcdeg}$ slices, 3 each in the
North and South Galactic Caps.  Completed in 1996, the LCRS contains
26,418 galaxy redshifts, of which 23,697 lie within the official
geometric and photometric limits of the survey.  Accurate $R$-band
photometry and sky positions for program objects were extracted from
CCD drift scans obtained on the Las Campanas Swope 1-m telescope;
spectroscopy was performed at the Las Campanas Du~Pont 2.5-m
telescope, originally via a 50-fiber Multi-Object Spectrograph (MOS),
and later via a 112-fiber MOS.  For observing efficiency, all the
fibers were used, but each MOS field was observed only once.  Hence,
the LCRS is a collection of 50-fiber fields (with nominal apparent
magnitude limits of $16.0 \leq R < 17.3$) and 112-fiber fields (with
nominal apparent magnitude limits of $15.0 \leq R < 17.7$); see
Figure~1.  Thus, selection criteria vary from field to field, but
these selection criteria are carefully documented and therefore easily
taken into account.  Observing each field only once, however, creates
an additional selection effect: the protective tubing of the
individual fibers prevents the spectroscopic observation of both
members of galaxy pairs within 55~arcsec of each other.  Hence, groups
and clusters can be undersampled, potentially causing physical groups
to be split by a ``friends-of-friends'' percolation algorithm and
resulting in the mis-estimate of general group properties.  We will
return to this problem in the next section.

In constructing the group catalogue, we have considered only those
LCRS galaxies within the official geometric and photometric borders of
the survey; we have furthermore limited this sample to galaxies having
redshifts in the range
\begin{equation}
7,500~\mbox{km~s}^{-1} \leq cz < 50,000~\mbox{km~s}^{-1}
\end{equation}
and luminosities in the range
\begin{equation}
-22.5 \leq M_R - 5\log h < -17.5 .
\end{equation}
To avoid group-member incompleteness at the extremal
distances of the sample, only groups within
\begin{equation}
10,000~\mbox{km~s}^{-1} \leq cz < 45,000~\mbox{km~s}^{-1}
\end{equation}
were admitted into the final group catalogue.  

[N.B.: Unless otherwise noted, all redshifts $z$ in this text are
corrected for motion relative to the dipole moment of the cosmic
microwave background (CMB; Lineweaver et al.\ 1996).]

\section{Extracting the Group Catalogue}

\subsection{The ``Friends-of-Friends'' Algorithm}

The LCRS group catalogue was extracted by means of an adaptive
``friends-of-friends'' percolation algorithm based upon that of HG82
and modified for use with comoving distances and field-to-field
sampling variations.

We outline the procedure as follows: First, a seed galaxy (``galaxy
$i$'') is selected which has not yet been classified as either a group
member or an isolated galaxy.  Every other non-classified galaxy $j$
in the survey sample is then tested to see if it lies within a
projected separation $D_{\rm L}$ and a velocity difference $V_{\rm L}$ of the seed
galaxy (note that both $D_{\rm L}$ and $V_{\rm L}$ are functions of both the field
$f$ and of the mean distance to galaxy pair $D_{\rm ave}$):
\begin{equation}
D_{ij} = 2 D_{\rm ave} \sin(\Theta_{ij}/2) \leq D_{\rm L}(D_{\rm ave}, f),
\end{equation}
where
\begin{equation}
D_{\rm ave} \equiv ( D(z_i) + D(z_j) ) / 2 ;
\end{equation}
and
\begin{equation}
V_{ij} = c \times | z_i - z_j | \leq V_{\rm L}(D_{\rm ave},f) .
\end{equation}
The distances $D(z)$ are comoving, 
\begin{equation}
D(z) = \frac{c}{H_0 q_0^2 (1+z)} [ q_0z + (q_0-1)( \sqrt{2 q_0 z + 1} - 1) ] 
\end{equation}
($q_0 = 0.5$ and $H_0 = h \times 100$~km~s$^{-1}$~Mpc$^{-1}$).  The
variable $\Theta_{ij}$ is the angular separation between the two
galaxies.  If no companions are found within $D_{\rm L}$ and $V_{\rm
L}$ of the seed galaxy, it is assigned ``isolated'' status and another
seed galaxy is sought.  If companions are found, they are added along
with the seed galaxy to a list of group members forming a new group.
In turn, the surroundings of each of these companions are combed for
the next level of ``friends.''  This loop is repeated until no further
companions are located, and the process is begun again by pursuing
another seed galaxy.  The group catalogue is complete only once every
galaxy in the redshift sample has been classified as either
``isolated'' or ``grouped.''  Only those groups containing three or
more members are included in the final catalogue.

The linking parameters, $D_{\rm L}$ and $V_{\rm L}$, are specified in
a manner which compensates for both the radial selection function and
the field-to-field sampling variations characteristic of the LCRS.
For each pair of galaxies,
\begin{equation}
D_{\rm L} = D_0 \times S_{\rm L} \hspace{1.0cm} \mbox{and} \hspace{1.0cm} V_{\rm L} = V_0 \times S_{\rm L}, 
\end{equation}
where $D_0$ and $V_0$ are $D_{\rm L}$ and $V_{\rm L}$, respectively,
for a given fiducial field at given fiducial redshift, and where
$S_{\rm L}$ is a linking scale which takes into account variations in
galaxy sampling rate.  It is defined by
\begin{equation}
S_{\rm L} \equiv \left[ \frac{ n^{\rm exp}(f,D_{\rm ave}) }{ n^{\rm exp}_{\rm fid}} \right]^{-1/3} ,
\end{equation}
where $n^{\rm exp}(f,D_{\rm ave})$ is the number density of galaxies
one would expect to observe at a comoving distance $D_{\rm ave}$ in
field $f$ for a randomly homogeneous distribution of galaxies having
the same selection function and sampling fraction as the LCRS redshift
catalogue; $n^{\rm exp}_{\rm fid}$ is $n^{\rm exp}(f,D_{\rm ave})$ for
a given fiducial field at a given fiducial redshift (Fig.~2).  Both
$n^{\rm exp}_{\rm fid}$ and $n^{\rm exp}(f,D_{\rm ave})$ are computed
by numerically integrating the field's selection function (see
Appendix~A).

To elaborate, the fiducial field is an idealized field with a given
set of characteristics.  The fiducial field is itself never generated.
It merely serves as the basis for the normalization of the linking
scale $S_{\rm L}$ in Equation~9.  For simplicity, we have chosen our
fiducial field to have 100\% sampling, flux limits of $15.0 \le R <
17.7$, and a Schechter (1976) luminosity function with the same parameter
values as the LCRS 112-fiber sample:
\begin{equation}
\alpha = -0.70, M^* = -20.29 + 5 \log h, \phi^* = 0.019h^3~\mbox{Mpc}^{-3}
\end{equation} 
(Lin et al.\ 1996).  Since it is roughly the median redshift of the
survey, we have chosen the fiducial redshift $cz_{\rm fid}$ to be
30,000~km~s$^{-1}$.

Moving on, we note that field-to-field sampling variations are of
particular concern when linking occurs across a field border.  This
concern is especially important in the case of a group situated on the
border between a 50-fiber field and a 112-fiber field, where a factor
of 2 discontinuity can occur in the expected surface density of
galaxies on the sky.  (Fortunately, only $\sim 2\%$ of the groups in
the final catalogue --- 28 out of 1495 --- straddle a 50/112 border.)
Therefore, when calculating the linking scale $S_{\rm L}$ for two
galaxies in two different fields, $n^{\rm exp}(f,D_{\rm ave})$ is
taken to be its average from the two fields,
\begin{equation}
n^{\rm exp}(f,D_{\rm ave}) \Longleftarrow (n^{\rm exp}(f_1,D_{\rm ave}) + n^{\rm exp}(f_2,D_{\rm ave})) /2 .
\end{equation}

But what of the artificial splitting of groups due to the LCRS's
55~arcsec fiber separation limit?  To avoid this problem, each of the
$\sim 1,000$ galaxies originally excluded from LCRS redshift catalogue
due to the fiber separation limit has been re-introduced into the
sample by assigning it a redshift equal to that of its nearest
neighbor convolved with a gaussian of width $\sigma = 200$~km~s$^{-1}$
(roughly the median line-of-sight velocity dispersion of a cleaned
LCRS group sample which excludes the 55-arcsec ``orphans'').  The
re-included galaxies subscribe to all the same photometric limits and
spatial borders that are imposed upon the original galaxy sample
(\S~2).

In closing, we note that, for a given choice of $D_0$ and $V_0$, this
algorithm leads to a unique group catalogue independent of the choice
of the original seed galaxy.  Due to the aforementioned field-to-field
variations in sampling, however, it is more intuitive to characterize
a group catalogue extracted from the LCRS not with $D_0$ for a certain
fiducial field, but with
\begin{equation}
\frac{\delta n}{n} = \frac{3}{ 4 \pi D_0^3 n^{\rm exp}_{\rm fid} } - 1 ,
\end{equation}
the corresponding number density enhancement of the surface contour
which delimits each group; the value $\delta n / n$ characterizes the
groups more generally, as it is valid no matter the field in which a
given group resides.

\subsection{The Choice of Linking Parameters}

So far, it has been shown how the superstructure of the group-finding
algorithm has been set into place.  The choice of values for $\delta n
/ n$ ($D_0$) and $V_0$, however, has yet to be presented and
explained.  We shall follow a course very similar to that of HG82 in
our justification of the two values ultimately adopted.

Take Figure~3 as a guide to the choice of $\delta n / n$ ($D_0$) and
$V_0$.  Their selection should satisfy a few basic criteria.  First,
the density enhancement contour sought should be high enough to limit
the number of interloper galaxies contained within a group, but not so
high that only the cores of rich clusters are found.  On the other
hand, for a thin-wedge geometry like that of an LCRS slice, the groups
selected should not be too loose; otherwise, edge effects become
excessive.  Furthermore, the value of $V_0$, as with the value of
$\delta n / n$ ($D_0$), should minimize the number of interlopers, but
without biasing group line-of-sight velocity dispersions toward
artificially low values.  As a definite upper limit, $V_0$ should not
exceed the radius (in~km~s$^{-1}$) of a typical void observed at
$cz_{\rm fid}$.  The range of acceptable values for $\delta n / n$
($D_0$) and $V_0$ are enclosed by the solid border in Figure~3. By
means of the following semi-quantitative arguments, it will be
concluded that the most reasonable values for the two linking
parameters are
\begin{equation}
\delta n / n = 80  \hspace{0.5cm} ( \Longleftrightarrow D_0 = 0.715~h^{-1}~\mbox{Mpc} ) 
\hspace{0.5cm} \mbox{and} \hspace{0.5cm} V_0 = 500\mbox{ km s}^{-1},
\end{equation}
which are denoted in Figure~3 by an asterisk. 	

First, consider the number of interlopers per galaxy, $n_{\rm I}$,
within a specified $D_{\rm L}$ and $V_{\rm L}$ of some galaxy.  For a
given galaxy, $n_{\rm I}$ can be roughly estimated by the equation
(HG82)
\begin{equation}
n_{\rm I} = \pi \left( \frac{D_{\rm L}}{D(z)} \right)^2 \left[
\frac{N_{V_{\rm L}}(cz)}{N_{\rm tot}} \right] \Sigma_{\rm gal} ,
\end{equation}
where $\Sigma_{\rm gal}$ is the surface number density of galaxies in
the redshift catalogue, $N_{\rm tot}$ is the total number of galaxies
in the sample, and $N_{V_{\rm L}}(cz)$ is the number of galaxies
within $V_{\rm L}$ of the galaxy's velocity $cz$.  (This measure is
actually an underestimate of $n_{\rm I}$, since it neglects the
correlation of galaxy positions on the sky, but, if we stringently
limit the number of interlopers per galaxy, it is adequate for our
purposes.)  The lower curve in Figure~3 denotes the locus $n_{\rm I} =
1$ for the the fiducial field at the fiducial redshift $cz_{\rm fid}
=$ 30,000~km~s$^{-1}$.  Below and to the right of this curve, where
$n_{\rm I} > 1$, the number of interlopers per galaxy is considered
excessive, and thus this curve constitutes one of our
boundaries. Figure~4 presents $n_{\rm I}$ for the redshift of each
galaxy in the LCRS sample for the values of $\delta n / n$ ($D_0$) and
$V_0$ listed in equation~13.  The bumps and dips in Figure~4 are due
to inhomogeneities within the distribution of galaxy velocities (the
wall-like structures) in the LCRS (Shectman et al.\ 1996).  Note that
the median number of interlopers per galaxy $n_{\rm I} \approx 0.2$
for our eventual choice of $\delta n / n$ ($D_0$) and $V_0$.  There is
only a slight large-scale trend evident in Figure~4, the fact of which
argues that the number of spurious groups in the final catalogue
should not be a strong function of redshift.

To preclude only finding the dense central regions of rich clusters,
an arbitrary upper limit to the density enhancement contour cut is set
at $\delta n/n = 200$, equivalent to assigning a value of
0.528~$h^{-1}$~Mpc to $D_0$.  In the quest for a catalogue of loose
groups, a lower density contrast cutoff is preferred.  Upon testing,
it was discovered that, below a contrast of $\delta n/n = 80$, edge
effects become a problem.  At these cutoffs, over half of the groups
must be excluded from the clean sample used in the study of group
properties; the group radii encroach upon the slice's borders.  Thus,
the density contrast $\delta n/n = 80$ was chosen for the group
catalogue, since it provides a reasonable compromise between the
pursuit of loose groups and the desire for a large clean sample.

Next, the value of $V_0$ should be chosen such that group velocity
dispersions are not overconstrained.  To avoid seriously
underestimating group velocity dispersions, we must set $V_0$ to a
value which accommodates the maximum {\em likely\/} physical velocity
dispersion, $\sigma^{\rm max,phys}_{\rm los}$.  Mathematically, the
maximum {\em possible\/} velocity dispersion of a group, $\sigma^{\rm
max,th}_{\rm los}$, is that obtained when the group is maximally
spread out in redshift --- i.e., when each group member is just within
the linking velocity $V_{\rm L}$ of its nearest neighbor:
\begin{equation}
\sigma^{\rm max,th}_{\rm los}(N_{\rm obs},V_{\rm L}) = 
       \frac{V_{\rm L}}{\sqrt{(N_{\rm obs} - 1)}} 
\left[ \sum_{i=1}^{N_{\rm obs}} \left( i - \frac{N_{\rm obs} + 1}{2} \right)^2 \right]^{\case{1}{2}} .
\end{equation}
For example, the maximum velocity dispersion possible for a group
containing $N_{\rm obs} = 10$ galaxies would be
\begin{equation}
\sigma^{\rm max,th}_{\rm los}(N_{\rm obs}=10,V_{\rm L}) \approx 3.03 V_{\rm L} .
\end{equation}
At the very least, we want this theoretical maximum {\em possible\/}
velocity dispersion to encompass --- i.e., to be greater than --- the
maximum {\em likely\/} physical velocity dispersion.  Due to the LCRS
galaxy selection function, a group containing $N_{\rm obs} = 10$
galaxies in a fiducial field at the fiducial velocity ($cz_{\rm fid} =
$30,000~km~s$^{-1}$) will typically be of Abell richness class $R
\approx 0$ (Abell 1958).  If we make the conservative assumption that
the maximum likely $\sigma_{\rm los}$ for a physical $R \approx 0$
``group'' is no more than about $1,200$~km~s$^{-1}$ (Zabludoff et al.\
1993; RPG97), we can set a lower limit for $V_0$ by means of the
relation
\begin{equation}
\sigma_{\rm los}^{\rm max,th}(N_{\rm obs}=10,cz=30,000~\mbox{km~s}^{-1}) \gtrsim 
\sigma_{\rm los}^{\rm max,phys} \approx 1,200\mbox{~km~s}^{-1},
\end{equation}
which yields $V_0 \gtrsim$ 400~km~s$^{-1}$.  This value provides the
leftmost boundary to the region of acceptable $(\delta n/n,V_0)$ in
Figure~3.  At the other extreme, the velocity linking parameter should
not be so large that galaxies are linked across a void diameter.
Thus, we give $V_0$ a maximum limit equal to the typical radius of
observed voids near $cz_{\rm fid} =$ 30,000~km~s$^{-1}$, or $V_0
\lesssim$ 2000~km~s$^{-1}$.  This maximum $V_0$ provides the rightmost
border in Figure~3.  To fine-tune $V_0$, the group algorithm was run,
using $\delta n/n = 80$, for four different values: $V_0 =$
500~km~s$^{-1}$, $V_0 =$ 1000~km~s$^{-1}$, $V_0 =$ 1500~km~s$^{-1}$,
and $V_0 =$ 2000~km~s$^{-1}$ (Fig.~5).  Many of the group velocity
dispersions are unaffected by the change in $V_0$, but a significant
number of groups, by adding progressively more outlying galaxies, see
a dramatic increase in $\sigma_{\rm los}$ as $V_0$ is increased from
500~km~s$^{-1}$ to 2000~km~s$^{-1}$.  In the end, $V_0 =$
500~km~s$^{-1}$ was chosen, due to its lower probability of
incorporating spurious groups into the catalogue, and due to its
ability to generate a catalogue with a relatively low
redshift-dependence in $\sigma_{\rm los}$.  This latter property
especially makes for a more homogeneous group sample.

\section{The Group Catalogue}

The full catalogue contains 1495 groups and includes 35\% of the LCRS
galaxy sample (Figs.~6, 7, \& 8).  The complete list of LCRS groups
and their individual properties is compiled in Table~1, which, due to
its size, is confined to the electronic version of this text.  Members
of the clean sample of 394 groups --- those which have barycenters
more than two pairwise separations [column (12)] from a slice edge,
crossing times [column (16)] less than a Hubble time, corrected
line-of-sight velocity dispersions [column (9)] greater than zero, and
no 55-arcsec ``orphan'' galaxies as group members --- are marked by an
asterisk in column (1).  Group properties are calculated according to
the prescriptions of RGH89, but modified for use with comoving
distances and field-to-field sampling variations.

Before we enter a description of the tabulated group properties, we
must note two caveats.  First, even the clean sample is unlikely to be
perfectly clean.  As noted, one of the rejection criteria used is the
removal of groups which are closer than two pairwise separations from
a slice edge.  In doing so, we have followed the lead of RGH89, who
used this same rejection criterion to define their clean group sample.
Clearly, having no ``edge-proximity'' rejection criterion would
include many groups which overflow the survey boundaries.  The
measured properties of these groups would be biased from their true
values due to their truncated membership.  Since the pairwise
separation, $R_{\rm p}$, is a measure of the group radius, excluding
groups which are closer than one $R_{\rm p}$ to a slice edge should go
far in counteracting this effect.  One must be careful, though, since
a truncated membership will both offset the position of the measured
group barycenter away from the slice edge and bias the measure of
$R_{\rm p}$ itself toward small values.  Thus, a group which in fact
extends over a slice border could still be accidentally included in
the clean sample.  Therefore, choosing an edge-proximity rejection
criterion of $2 \times R_{\rm p}$ --- although still not perfect ---
is much safer.  One could even make more stringent demands, requiring
groups in the clean sample to be at least three or four times their
measured $R_{\rm p}$ from a slice edge.  Here, however, one must worry
about biasing the clean sample unnecessarily towards only the most
compact systems.  In Table~2, we see the effects of steadily
increasing the threshold of edge-proximity rejection on the resulting
clean samples: although the sample line-of-sight velocity dispersion,
$\sigma_{\rm los}$, is not strongly affected, the sample pairwise
separation and harmonic radius [$R_{\rm h}$; see column (14) of
Table~1] both drop precipitously for these larger values of the
rejection criterion.  Therefore, as a compromise, we have chosen a
value of $2 \times R_{\rm p}$.

Our second caveat is that the clean sample is not necessarily the best
sample to use for all purposes.  Since the clean sample excludes
groups containing 55-arcsec ``orphans'', on average this sample
discriminates against groups with dense cores.  On some occasions (as
in \S~6), a superset of the clean sample -- one which does not exclude
groups merely because they harbor an ``orphan'' -- is the preferred
catalogue.  The clean sample, as defined, however, is the most
conservative catalogue -- the sample with the fewest extrinsic
assumptions attached to it.  Therefore, in matters of discussing
intrinsic mean group properties, we shall use this clean group sample.

With these caveats in mind, the columns in Table~1 are as follows:

\noindent Column (1): A running group identification number, $N_{\rm
grp}$, for a given LCRS slice.  The slice declination and $N_{\rm
grp}$ form the basis of the IAU-registered naming convention for LCRS
loose groups, which is of the form
\begin{displaymath}
\mbox{LCLG} \; -\!D\!D \; N\!N\!N ,
\end{displaymath}
where LCLG stands for ``Las Campanas Loose Group,'' $-\!D\!D$ is
the (zero-padded) declination for the LCRS slice wherein the group
resides, and $N\!N\!N$ is the (zero-padded) $N_{\rm grp}$.  For more
information, see the online Dictionary of Nomenclature of Celestial
Objects ({\tt http://cdsweb.u-strasbg.fr/cgi-bin/Dic}).

\noindent An asterisk appended to $N_{\rm grp}$ in Table~1 indicates
that the group is a member of the clean sample.

\noindent Columns (2 -- 4): A weighted measure of the B1950.0 right
ascension (in HH~MM~SS.ss format) of the group's barycenter,
\begin{equation}
\alpha_{1950.0} = \frac{\sum_{i=1}^{N_{\rm obs}} w_i \alpha_i}{\sum_{i=1}^{N_{\rm obs}} w_i}, 
\end{equation}
where 
\begin{equation}
w_i \equiv \frac{1}{n^{\rm exp}(f_i,D(z_i))}
\end{equation}
This weighting factor --- which is proportional to the inverse of the
selection function --- helps to counteract a bias resulting from a
group straddling two fields with different galaxy sampling
characteristics; a discussion on how $n^{\rm exp}(f_i,D(z_i))$ is
estimated can be found in Appendix~A.  $N_{\rm obs}$, the number of
observed group members, is listed in column (11).

\noindent Column (5-7): A weighted measure of the B1950.0 declination 
(in sDD~MM~SS.s format) of the group's barycenter,
\begin{equation}
\delta_{1950.0} = \frac{\sum_{i=1}^{N_{\rm obs}} w_i \delta_i}{\sum_{i=1}^{N_{\rm obs}} w_i}, 
\end{equation}
where $w_i$ is as defined in equation~19. $N_{\rm obs}$ is listed in
column (11).

\noindent Column (8): The group's redshift, $z_{\rm cmb}$, with
respect to the local comoving frame.  It is taken as the (unweighted)
mean of the members' redshifts. (See Figs.~9 \& 10.)

\noindent Column (9): The group line-of-sight velocity dispersion, in
km~s$^{-1}$, corrected for relativistic effects (Harrision 1974),
\begin{equation}
\sigma_{\rm los} = \frac{1}{1 + <z>} \sqrt{ 
\frac{\sum_{i=1}^{N_{\rm obs}} (cz_i - <cz>)^2}{(N_{\rm obs} - 1)} } 
\end{equation}
(Fig.~11).  The random errors in the LCRS galaxy redshift
measurments are also removed, in quadrature, assuming $\sigma_{cz} =
67$~km~s$^{-1}$ (Shectman et al.\ 1996):
\begin{equation}
\sigma_{\rm los}  \Longleftarrow 
\cases{ 
\sqrt{ \sigma_{\rm los}^2 - (67~{\rm km~s}^{-1})^2 }, & if $\sigma_{\rm los} > 67~{\rm km~s}^{-1}$ ; \cr
0, & otherwise. \cr
}
\end{equation}
$N_{\rm obs}$ is listed in column (11).  

\noindent Column (10): A formal estimate of the standard error in
$\sigma_{\rm los}$, in km~s$^{-1}$.  If a normal distribution is
assumed for the line-of-sight velocities, the standard error for
$\sigma_{\rm los}^2$ can be expressed as
\begin{equation}
\sigma(\sigma_{\rm los}^2) = \sigma_{\rm los}^2 \sqrt{ \frac{2}{N_{\rm obs}} 
\left( 1 - \frac{1}{N_{\rm obs}} \right) }
\end{equation}
(Deming 1950).  By means of propagation of errors (Bevington 1969),
this expression yields
\begin{equation}
\sigma(\sigma_{\rm los}) = \frac{1}{2} \sigma_{\rm los} \sqrt{ \frac{2}{N_{\rm obs}} 
\left( 1-\frac{1}{N_{\rm obs}} \right) },
\end{equation}
which is employed for the estimates listed in Column (10).  $N_{\rm obs}$ 
is listed in column (11).

\noindent Column (11): $N_{\rm obs}$, the observed number of LCRS
galaxies (including any 55-arcsec ``orphans'') comprising the group
(Fig.~12).  Included in $N_{\rm obs}$ are only those LCRS galaxies
which the lie within the official geometric and photometric borders of
the survey and which subscribe to the redshift and absolute magntitude
limits set forth in equations~1 and 2.

\noindent As is typical for ``friends-of-friends'' group catalogues, the
distribution of $N_{\rm obs}$ for the LCRS group catalogue is heavily
skewed toward small values: the median $N_{\rm obs}$ is 3.

\noindent Column (12):  The mean pairwise separation,
\begin{equation}
R_{\rm p} = \frac{8 D_{\rm grp}}{\pi} 
\sin \left[ \frac{1}{2} <\theta_{ij}> \right] ,
\end{equation}
where 
\begin{equation}
<\theta_{ij}> \equiv \frac{\sum_{i} \sum_{j>i} w_i w_j \theta_{ij}}{\sum_{i} \sum_{j>i} w_i w_j} , 
\end{equation}
and where 
$D_{\rm grp}$ is the comoving distance to the group,
$\theta_{ij}$ is the angular separation between group members $i$ and
$j$, and $w_i$ and $w_j$ are the respective weights for
$i$ and $j$ (equation~19).  $R_{\rm p}$ has dimensions $h^{-1}$~Mpc.
See Figure~13.

\noindent Column (13):  An estimate of the rms error in $R_{\rm p}$,
\begin{equation}
\sigma_{R_{\rm p}} = \left( \frac{4}{\pi} \right) D_{\rm grp} \sigma_{<\theta_{ij}>}, 
\end{equation}
where 
\begin{equation}
\sigma_{<\theta_{ij}>} = \sqrt{
\frac{ N_{\rm pair} \sum_{i} \sum_{j>i} \left( w_i w_j \theta_{ij} \right)^2 - 
       \left( \sum_{i} \sum_{j>i} w_i w_j \theta_{ij}  \right)^2 }{
       \left( N_{\rm pair} - 1 \right) \left( \sum_{i} \sum_{j>i} w_i w_j \right)^2 } 
} ,
\end{equation}
where $N_{\rm pair}$ is the number of distinct galaxy pairs in the
group, and where $D_{\rm grp}$, $\theta_{ij}$, $w_i$, and $w_j$ are as
described for $R_{\rm p}$ (equations~25 and 26).  

\noindent Equation~27 was derived from Equation 25, assuming that $\sin
(0.5<$$\theta_{ij}$$>) \approx 0.5<$$\theta_{ij}$$>$ for the typical
group angular sizes encountered in this group catalogue, and that the
contribution to $\sigma_{R_{\rm p}}$ by the rms error in $D_{\rm grp}$
is insignificant compared to the contribution by $<$$\theta_{ij}$$>$.
Equation~28 was derived from equation~26 via a straightforward (albeit
tedious) application of propagation of errors.

\noindent The units for $\sigma_{R_{\rm p}}$ are $h^{-1}$~Mpc.

\noindent Column (14):  The harmonic radius,
\begin{equation}
R_{\rm h} = \pi D_{\rm grp} 
\sin \left[ \frac{1}{2} <\theta_{ij}^{-1}>^{-1} \right] , 
\end{equation}
where 
\begin{equation}
<\theta_{ij}^{-1}> \equiv \frac{\sum_{i} \sum_{j>i} w_i w_j \theta_{ij}^{-1}}{\sum_{i} \sum_{j>i} w_i w_j} .
\end{equation}
$D_{\rm grp}$, $\theta_{ij}$, and $w_i$ and $w_j$ are as described for
$R_{\rm p}$ (equations~25 and 26).  $R_{\rm h}$ has dimensions
$h^{-1}$~Mpc.  See Figure~14.

\noindent Column (15):  An estimate of the rms error in $R_{\rm h}$, 
\begin{equation}
\sigma_{R_{\rm h}} = \left( \frac{\pi}{2} \right) 
                     \left( \frac{D_{\rm grp}}{<\theta_{ij}^{-1}>^2} \right)
                      \sigma_{<\theta_{ij}^{-1}>} , 
\end{equation}
where
\begin{equation}
\sigma_{<\theta_{ij}^{-1}>} = \sqrt{
\frac{ N_{\rm pair} \sum_{i} \sum_{j>i} \left( w_i w_j \theta_{ij}^{-1} \right)^2 - 
       \left( \sum_{i} \sum_{j>i} w_i w_j \theta_{ij}^{-1}  \right)^2 }{
       \left( N_{\rm pair} - 1 \right) \left( \sum_{i} \sum_{j>i} w_i w_j \right)^2 } 
} .
\end{equation}
$N_{\rm pair}$ is the number of distinct galaxy pairs in the group;
$D_{\rm grp}$, $\theta_{ij}$, $w_i$, and $w_j$ are as described for
$R_{\rm p}$ (equations~25 and 26).

\noindent The derivation of equations~31 and 32 closely mimics that of
equations~27 and 28.

\noindent The units for $\sigma_{R_{\rm h}}$ are $h^{-1}$~Mpc.

\noindent Column (16):  The crossing time for the group,
\begin{equation}
t_{\rm cr} = \frac{3}{5^{3/2}} \frac{R_{\rm h}}{\sigma_{\rm los}} ,
\end{equation}
in units of the Hubble time ($H_0^{-1}$); see Figure~15.  This measure
is heavily influenced by the relative values of the linking parameters
$D_{\rm 0}$ (which determines $R_{\rm h}$) and $V_{\rm 0}$ (which
determines $\sigma_{\rm los}$).  Following Gott \& Turner (1977), it
can be estimated that the time for a uniform sphere to undergo
complete virialization is
\begin{equation}
t_{\rm vir} \sim 3 \pi t_{\rm cr}.
\end{equation}
Hence, all groups with $t_{\rm cr} \lesssim 0.11H_0^{-1}$ should have
had enough time to virialize completely within the age of the Universe
[but see Diaferio et al.\ (1993)].  Thus, it can be deduced that
roughly half of the LCRS clean sample could have undergone
complete virialization.

\noindent Column (17):  An estimate of the rms error in $t_{\rm cr}$,
\begin{equation}
\sigma_{t_{\rm cr}} = t_{\rm cr} 
\sqrt{ \frac{\sigma_{R_{\rm h}}^2}{R_{\rm h}^2} + 
       \frac{\sigma_{\sigma_{\rm los}}^2}{\sigma_{\rm los}^2} } ,
\end{equation}
in units of the Hubble time ($H_0^{-1}$); $\sigma_{\rm los}$,
$\sigma_{\sigma_{\rm los}}$, $R_{\rm h}$, and $\sigma_{R_{\rm h}}$
come from columns~(9), (10), (14), and (15), respectively.
Equation~35 was derived by standard propagation of errors analysis of
equation~33.  

\noindent Column (18):  The group's virial mass,
\begin{equation}
M_{\rm vir} = \frac{6 \sigma_{\rm los}^2 R_{\rm h}}{G} ,
\end{equation}
where $G$ is the gravitational constant (Fig.~16).  This estimate
assumes that the groups are virialized and that the galaxies trace the
mass distribution within the group [see, for example, Binney \&
Tremaine (1987), Chapter 10, Section 2.3].  $M_{\rm vir}$ is in units
of $h^{-1}~M_{\sun}$.

\noindent Column (19): An estimate of the rms error in $M_{\rm vir}$,
\begin{equation}
\sigma_{M_{\rm vir}} = M_{\rm vir} 
\sqrt{ \frac{4 \sigma_{\sigma_{\rm los}}^2}{\sigma_{\rm los}^2} + 
       \frac{\sigma_{R_{\rm h}}^2}{R_{\rm h}^2} } , 
\end{equation}
where $\sigma_{\rm los}$, $\sigma_{\sigma_{\rm los}}$, $R_{\rm h}$,
and $\sigma_{R_{\rm h}}$ come from columns~(9), (10), (14), and (15),
respectively.  Equation~37 was derived by propagation of errors
analysis of equation~36.  Note that the large random errors inherent
to $\sigma_{\rm los}$ and to $R_{\rm h}$ propagate into estimates for
$M_{\rm vir}$.  The units for $\sigma_{M_{\rm vir}}$ are $h^{-1}~M_{\sun}$.

\noindent Column (20): The total group luminosity in the LCRS
$R$-band, $L_{\rm tot}$, corrected for selection effects to account
for galaxies not observed by the LCRS (Fig.~17); $L_{\rm tot}$ is in
units of solar luminosity ($h^{-2}~L_{\sun}$), in which the $R$-band
absolute magnitude of the Sun is taken to be $M_{R;\sun} = +4.52$
(Pinsonnealt 1992).  The mathematical apparatus behind the correction
factor can be found in Appendix~B.

\noindent Column (21):  An estimate of the rms error in $L_{\rm tot}$, 
$\sigma_{L_{\rm tot}}$, obtained by summing the rms errors of the
individual components of $L_{\rm tot}$ in quadrature.  Details can be
found in Appendix~B.  The units for $\sigma_{L_{\rm tot}}$ are
$h^{-2}~L_{\sun}$ (LCRS $R$-band).

\noindent Column (22): The ratio, $L_{\rm rat}$, by which the sum of
the luminosities of the observed $N_{\rm obs}$ galaxies must be
multiplied in order to obtain an estimate of the group's total (LCRS
$R$-band) luminosity, $L_{\rm tot}$, 
\begin{equation}
L_{\rm rat} \equiv \frac{L_{\rm tot}}{\sum_{i=1}^{N_{\rm obs}} L_i}
\end{equation}
Over the clean sample, the median $L_{\rm rat}$ is $\approx 5.4$
for the 50-fiber groups and $\approx 2.4$ for the 112-fiber groups.

\noindent Column (23): The group mass-to-light ratio in the LCRS
$R$-band, $M/L$, in units of $h~M_{\sun}/L_{\sun}$ (Fig.~18).  For
comparison with the mass-to-light ratios for groups from other
redshift catalogues (in particular, those based upon the
de~Vaucouleurs $B(0)$-band), it is convenient to note that
\begin{equation}
M/L_{B(0)} \sim 1.1 M/L \mbox{ (LCRS)} .
\end{equation}
The uncertainties in $M_{\rm vir}$ and in $L_{\rm tot}$ tend to give
large errors for the mass-to-light ratios of individual groups.

\noindent Column (24):  An estimate of the rms error in $M/L$,
\begin{equation}
\sigma_{M/L} = \left( \frac{M}{L} \right)
\sqrt{ \frac{\sigma_{M_{\rm vir}}^2}{M_{\rm vir}^2} + 
       \frac{\sigma_{L_{\rm tot}}^2}{L_{\rm tot}^2} }
\end{equation}
where $M_{\rm vir}$, $\sigma_{M_{\rm vir}}$, $L_{\rm tot}$, and
$\sigma_{L_{\rm tot}}$ come from columns~(18), (19), (20), and (21),
respectively.  Equation~40 was obtained by means of propagation of
errors analysis of equation~39.  The units for $\sigma_{M/L}$ are
$h~M_{\sun}/L_{\sun}$.

\noindent Column (25): An estimate of the group's Abell counts,
$C_{\rm grp}$ (Fig.~19).  Abell (1958) defined his counts to be the
number of galaxies, corrected for background contamination, in the
magnitude interval $m_3$ to $m_3 + 2$, that lie within a
1.5~$h^{-1}$~Mpc projected separation of a cluster's center; the
magnitude $m_3$ is the magnitude of the third brightest cluster
member.  Due to the sampling characteristics and the relatively small
range of apparent magnitudes within the LCRS, we could not use Abell's
definition directly.  We had instead to derive each group's Abell
counts via a Schechter function; the details of the method can be
found in Appendix~C.  When compared with actual Abell clusters within
the LCRS volume (\S~6), we find the following relation between the
LCRS group counts estimate, $C_{\rm grp}$, and the revised Abell
cluster values given by ACO:
\begin{equation}
C_{\rm grp} \sim 0.19C_{\rm ACO} + 12 .
\end{equation}

\noindent Column (26): An estimate of the error in the Abell counts
$C_{\rm grp}$, based upon Poisson statistics of the observed number of
LCRS member galaxies within a projected distance of 1.5~$h^{-1}$~Mpc
of the group center ($N_{\rm obs}^{1.5}$), 
\begin{equation}
\sigma_{C_{\rm grp}} \sim \frac{C_{\rm grp}}{ \sqrt{ N_{\rm obs}^{1.5} } } .
\end{equation}

\noindent Column (27): The type of group --- one within the borders of
a 50-fiber field (Type~1), one within the borders of a 112-fiber field
(Type~2), or one straddling the border of a 50- and a 112-fiber field
(Type~3).

\noindent Column (28): A column to reference any applicable notes for the
given group.  These table notes are as follows:

\begin{description}

\item[a:] the group's crossing time $t_{\rm cr}$ is greater than a
          Hubble time ($H_0^{-1}$).

\item[b:] the group's barycenter is closer than $2R_{\rm p}$ to a
          slice edge.

\item[c:] the group contains at least one 55~arcsec ``orphan'' with a
          mock redshift.

\item[d:] the group's line-of-sight velocity dispersion, corrected for
          galaxy velocity errors, is less than or equal to 0~km~s$^{-1}$.

\end{description}

Table~3 lists the members of each group.  Due to its great size (10761 
lines), it is provided only in electronic format. Each group is
introduced by a header composed of its group number designation
$N_{\rm grp}$, its redshift $z_{\rm cmb}$, and its line-of-sight
velocity dispersion $\sigma_{\rm los}$ [Columns (1), (8), and (9) of
Table~1, respectively].  The columns for Table 2 are as follows:

\noindent Column (1): A group member identification number.

\noindent Columns (2-4): The group member's right ascension in 1950.0
coordinates.

\noindent Columns (5-7): The group member's declination in 1950.0
coordinates.

\noindent Column (8):  The group member's LCRS $R$-band isophotal magnitude.

\noindent Column (9):  The group member's LCRS $R$-band central magnitude.

\noindent Column (10): The group member's heliocentric velocity,
$cz_{\rm helio}$, in km~s$^{-1}$.

\noindent Column (11): The group member's spectrum type (e $=$
contains strong emission lines; c = a continuum with absorption lines;
b $=$ contains both absorption lines and moderate emission lines; m
$=$ mock velocity --- i.e., the galaxy was one of the 55-arcsec
``orphans'' excluded from the spectroscopic survey due to its
proximity to another galaxy).

The individual group member properties are derived directly from a
working copy of the published LCRS galaxy catalogue; further details
can be found in Shectman et al.\ (1996).

\section{Comparison of Various Group Catalogues}

Table~4 contains information regarding the general properties of
groups both in the LCRS catalogue and from other catalogues.  The
values tabulated for the eight non-LCRS group catalogues have been
taken from the original papers, and, where necessary, these values
were converted to be consistent with the property definitions detailed
in \S~4.  The median properties for the LCRS catalogue were derived
from the clean sample of 394 groups; the fraction of grouped
galaxies, from the full sample of 1495 groups.

\subsection{LCRS Group Catalogues}

We have broken the LCRS groups catalogue into three sub-catalogues for
the purpose of inter-comparison; the sub-catalogues include the LCRS
groups from 50-fiber fields, the LCRS groups from 112-fiber fields,
and the LCRS groups which straddle a 50/112 border.  Furthermore, we
can compare the present LCRS group catalogue with a precursor based
upon the LCRS $-6\arcdeg$ slice (Tucker 1994; henceforth, T94).  The
properties of these LCRS group samples are tabulated under the LCRS
main heading in Table~4.

Note that there is substantial variation in group properties among the
three sub-catalogues.  These variations are also apparent in Figures
11 -- 19.  First, we can discount the oddities apparent in the 50/112
group properties due to poor number statistics (there are only 5 of
these groups in the clean sample) and due to the difficulties of
linking group members across a 50/112 border.  In any case, if not for
its aberrantly large $\sigma_{\rm los}$'s --- which in turn affect
estimates of $M_{\rm vir}$ and $M/L$ --- the 50/112 sample properties
would closely match those of the 50-fiber sample.  Unfortunately, the
discrepancies between the 50- and the 112-fiber samples' group
properties are somewhat harder to dismiss.  Great effort was put forth
into accounting for field-to-field sampling variations in the
group-finding algorithm, including using appropriate luminosity
functions and surface brightness cutoff functions for each of the
northern 50-fiber, the southern 50-fiber, the northern 112-fiber, and
the southern 112-fiber field types (Lin et al. 1996).  Apparently,
some small residual selection bias between the 50- and 112-fields
remains.  Fortunately, we can still follow one of two options: (1) we
can choose to ignore the 50-fiber groups altogether (they make up only
$\approx 20\%$ of the total number of groups), or (2) we can note that
both the 50-fiber and the 112-fiber group properties fall within the
general range of typical group properties from other surveys (\S~5.2)
and thus consider the combined LCRS group catalogue as being
representative of groups as a whole.  Unless otherwise noted, we will
take the latter course in the following sections.

Finally, consider the properties from an earlier version of the LCRS
group catalogue (T94).  This group catalogue is composed of only
50-fiber data from the LCRS $-6\arcdeg$ slice, so it is not surprising
that many of its general properties have values which approximate
those of the present 50-fiber sample.  One major difference is the
percentage of galaxies in groups.  The relatively small value for the
T94 sample may be attributed at least in part to the fact that the
earlier LCRS group-finding algorithm ignored 55-arcsec ``orphans,''
thus disrupting many groups into doublets or isolated galaxies which
were excluded from the resulting T94 catalogue.

\subsection{Group Catalogues from Other Redshift Surveys}

It is instructive to compare the LCRS group catalogue with those
derived from other galaxy redshift surveys.  In this section we shall
look at nine other group catalogues extracted from five different
surveys and relate their properties to those of LCRS groups.  The nine
group catalogues considered are those by GH83, NW87, N93, MFW93,
MdCL89, RGH89, RPG97, TB98, and RZZ99.  The five surveys are the CfA1
(Huchra et al.\ 1983), the SSRS (da Costa et al.\ 1988,1991), the CfA2
(de~Lapparent, Geller, \& Huchra 1988; Huchra et al.\ 1990; Huchra,
Geller, \& Corwin 1995), the PPS (Giovannelli \& Haynes 1993; Wegner,
Haynes, \& Giovanelli 1993), and the ESP (Vettolani et al.\ 1997).
Table~5 summarizes the general properties of the different survey
samples used to generate these group catalogues; for comparison, the
characteristics of the LCRS sample used in this paper are also
included (note that the LCRS sample contains substantially more
galaxies and encloses the largest volume of any of the survey samples
listed).  The interested reader is urged to consult the original
papers for details.

Now, as one might suspect from the small-but-statistically-significant
differences between subsamples of the LCRS group catalogue (\S~5.1;
Figs.~11 -- 19), it is unlikely that another group catalogue ---
extracted from a different survey with different galaxy selection
criteria using another variation of a ``friends-of-friends''
group-finding algorithm --- would be from the same statistical parent
population of groups as the LCRS group catalogue.  To verify this, we
have used the Kolmogorov-Smirnov (KS) Test to compare the physical
properties of the LCRS groups with those of the five group catalogues
available in machine-readable form (GH83, MdCL89, RPG97, TB98, and
RZZ99).  The results of our analysis are shown in Figures~21 -- 27 and
summarized in Table~6.  Notice that most of the formal KS
probabilities are so low that we needed to list them in logarithmic
form in order to make Table~6 readable.  Thus, these group catalogues
are not extracted from the same parent population as the LCRS groups:
their properties, although roughly similar (Table~4), do differ
significantly.

There is a hidden benefit in these very low KS probabilities.  The
other four group catalogues are not available in machine-readable form
(and two of those were not even published in paper form).  Since these
four have roughly the same number statistics and variations in
physical properties as the other five, we can with some safety assume
that the physical properties of the groups from these four catalogues
also differ significantly from those of the LCRS group sample.  We can
therefore exploit the one statistic that is published for the
properties of all nine of these group catalogues --- the median --- as
a reasonable measure of comparison.

Now, in comparison with these other group catalogues, it is clear from
Table~4 that the median properties of the full LCRS group catalogue
(those under the ``All'' heading) are fairly typical.  This is
comforting, since it shows that all these group catalogues are looking
at roughly the same sort of systems.  But can more be said?  Due to
differing survey characteristics and differing group-extracting
parameters, it is notoriously difficult to make cross-catalogue
comparisons.  Nonetheless, rough comparisons can be made.

Of the properties that Table~4 lists, two are primarily $D_{\rm
L}$-dependent ($<$$R_{\rm h}$$>_{\rm med}$, $<$$R_{\rm p}$$>_{\rm med}$), one
is primarily $V_{\rm L}$-dependant ($<$$\sigma_{los}$$>_{\rm med}$), and
several are dependent in more-or-less complex ways on both $D_{\rm L}$
and $V_{\rm L}$ (\# of groups in catalogue, \# of groups in
clean sample, \% of galaxies in groups, $<$$H_0 t_{\rm cr}$$>_{\rm
med}$, $<$$M_{\rm vir}$$>_{\rm med}$, $<$$L_{\rm tot}$$>_{\rm med}$,
$<$$M/L$$>_{\rm med}$).  Let us designate the properties which depend
primarily only on $D_{\rm L}$ {\em or\/} $V_{\rm L}$ as {\em simply
derived\/} properties, and those that depend more complexly on both
$D_{\rm L}$ {\em and\/} $V_{\rm L}$ as {\em complexly derived\/}
properties.  Of the simply derived properties, those depending mainly
on $D_{\rm L}$ can be considered density defining quantities; the one
that mainly depends on $V_{\rm L}$ ($<$$\sigma_{los}$$>_{\rm med}$)
defines the mean gravitational energy content of the systems.  Many of
the complexly derived properties are straightforward functions of
combinations of the simply derived properties (see \S~4).

Let us first consider the simply derived properties.  

What can we say about the relative mean densities of the groups
from the different catalogues?  Consider $<$$R_{\rm h}$$>_{\rm med}$.
For the catalogues which on average contain the denser groups,
$<$$R_{\rm h}$$>_{\rm med}$ should be relatively small.  Under this
criterion, we find that the TB98 and RPG97 catalogues contain the
densest systems on average; the LCRS, GH83, MdCL89, RGH89, and N93
catalogues, those of average density; and the NW87 catalogue, the
least dense.  $<$$R_{\rm p}$$>_{\rm med}$ is another indicator of
relative group density, and, albeit with some changes in rank, roughly
the same trend is seen.  Thus, we can conclude that the TB98 and RPG97
groups are the densest on average; the LCRS, GH83, MdCL89, RGH89, and
N93 groups are average systems; and the NW87 groups are the least
dense.

Next, what can we infer about the mean gravitational energy content of
the various catalogues --- i.e, which are the ``hottest'' systems, the
systems with the highest $<$$\sigma_{los}$$>_{\rm med}$?  From
Table~4, we see that the GH83 and RGH89 catalogues have high
$<$$\sigma_{los}$$>_{\rm med}$, the LCRS, MdCL89, RPG97, TB98, and
RZZ99 catalogues have intermediate values, and the NW87, N93, and
MFW93 catalogues are the ``coolest'' of the systems listed (this is a
reflection in part on the functional forms of the linking lengths used
in NW87, N93, and MFW93, which strongly limit velocity outliers).

The values for the complexly derived quantities are by their very
natures less certain.  We will consider three specifically ---
$<$$M_{\rm vir}$$>_{\rm med}$, $<$$H_0 t_{\rm cr}$$>_{\rm med}$, and
$<$$M/L$$>_{\rm med}$.

Consider $<$$M_{\rm vir}$$>_{\rm med}$: the GH83 and the RGH89 systems
are the most massive, followed in order of decreasing mass by the
LCRS, the RPG97, the MdCL89, the NW87, and the TB98 systems.  It may
seem strange that the catalogue containing the densest systems (TB98)
is also the catalogue with the least massive systems.  In fact, this
is a consequence of the relatively small linking lengths $D_{\rm L}$
that TB98 used to extract his groups from the PPS: his groups do not
extend out to the same radii as those in the ``looser'' catalogues, so
they enclose less mass.  On the other hand, his groups are the most
virialized systems in Table~4 (based upon their short crossing times),
so his catalogue estimate for $<$$M_{\rm vir}$$>$ is probably the most
accurate.  (The LCRS groups catalogue have a somewhat longer median
crossing time, but one can still expect a large fraction of them to be
virialized --- see equation~34.)

We can also ask which group catalogues give evidence for the most (or
least) dark matter.  The LCRS group catalogue has a $<$$M/L$$>_{\rm
med}$ of $171~h~M_{\sun}/L_{\sun}$ in the LCRS $R$-band; if we convert
this to the de~Vaucouleurs $B(0)$-band (equation~39) used by the other
group catalogues with measured mass-to-light ratios, we find
$<$$M/L$$>_{\rm med} = 188~h~M_{\sun}/L_{\sun}$.  This is fairly
average.  Of the other group catalogues with measured values of
$<$$M/L$$>_{\rm med}$, the N93 and RGH89 catalogues show lower
mass-to-light ratios, and the RPG97 and NW87 show higher.

In conclusion, we can state the LCRS group catalogue is quite average
in its simply derived properties --- its groups are of average density
and gravitional energy content.  Furthermore, it is not particularly
distinct in its complexly derived properties --- the groups being
moderately massive with relatively long crossing times and average
mass-to-light ratios.  It is, however, a very large catalogue of
groups in a wide variety of environments.  The LCRS values for the
properties tabulated here are among the least-biased presently
available.

\section{Abell Clusters in the LCRS Group Catalogue}

The original Abell Catalogue (Abell 1958) is comprised of 2712 rich,
highly dense clusters in the northern sky.  Revised and expanded to
include southern clusters, an updated Abell Catalogue (ACO) now
contains 4073 rich clusters over a significant fraction of the whole
sky. Each cluster contains within a projected radius of $1.7~{\rm
arcmin}/z$ ( $\approx 1.5h^{-1}$~Mpc) from its center at least 30
member galaxies in the magnitude range $m_3$ to $m_3+2$, where $m_3$
is the apparent magnitude of the third brightest cluster member.  The
Abell/ACO catalogue nominally encompasses a redshift range of $0.02
\lesssim z \lesssim 0.20$; the redshifts are estimated empirically
based upon the apparent magnitude of the tenth brightest member,
$m_{10}$, and tend to be accurate to within a factor of 2.  Each
cluster is classified according to distance class $D$, estimated from
$m_{10}$, and according to richness class $R$, based on the number of
members meeting the above criteria for projected radial distance and
apparent brightness.  Furthermore, a supplementary catalogue of
southern clusters too poor or too distant to be included in the main
catalogue contains an additional 1174 clusters (Abell S001 -- S1174).

How many Abell clusters do we expect to identify within the LCRS
group catalogue?  There are 206 Abell clusters (including those from
the supplementary catalogue) within the sky-projected confines of the
six LCRS slices.  If we confine ourselves to a superset of the
clean sample --- one which includes groups with 55~arcsec
``orphans'' meeting the other criteria for inclusion into the
clean sample --- there are 735 LCRS loose groups which could be
matched.  The Abell catalogue is purported to be complete to a
redshift of 0.20, but the LCRS group catalogue only goes to a redshift
of 0.15; so we expect $\sim (0.15/0.20)^3 \sim \onehalf$ of Abell
clusters to be lost due to distance.  Furthermore, if we assume a
typical pairwise separation of $R_{\rm p} \approx 0.6h^{-1}$~Mpc, the
effective volume of the LCRS slices for the superset sample is
again reduced by about half.  Therefore, we expect about $\onehalf
\times \onehalf \times 206 \approx 50$ matches between Abell clusters
and LCRS groups.

We find 54 matches, which are described in Table~7.  The columns are
as follows:

\noindent Column (1): The LCRS group's running identification number
for the given slice ($N_{\rm grp}$).

\noindent Column (2): The LCRS group's $\alpha$ in equinox 1950.0
coordinates ($\alpha_{\rm grp}$).

\noindent Column (3): The LCRS group's $\delta$ in equinox 1950.0
coordinates ($\delta_{\rm grp}$).

\noindent Column (4): The LCRS group's redshift ($z_{\rm grp}$).

\noindent Column (5): The observed number of galaxies from the LCRS
official spectroscopic sample that lie within the group ($N_{\rm
obs}$).

\noindent Column (6): An estimate of the Abell Richness for that group
($C_{\rm grp}$).

\noindent Column (7): The name of the Abell cluster match to this 
group.  

\noindent Column (8): The Abell cluster's $\alpha$ in equinox 1950.0
coordinates ($\alpha_{\rm ACO}$).

\noindent Column (9): The Abell cluster's $\delta$ in equinox 1950.0
coordinates ($\delta_{\rm ACO}$).

\noindent Column (10): The Abell cluster's redshift ($z_{\rm ACO}$), if
known.

\noindent Column (11): The Abell cluster's richness class ($R$).

\noindent Column (12): The Abell cluster's distance class ($D$).

\noindent Column (13): The Abell cluster's Abell counts, as measured
by ACO ($C_{\rm ACO}$).

\noindent Column (14): The angular separation in arcminutes between
the measured group center and the Abell cluster center as reported by
ACO.  A group-cluster pair was considered a ``match'' if this angular
separation was less than 12~arcmin ($\sim 1~h^{-1}$~Mpc at the median
redshift of the LCRS).

Of these matches, we see that some are quite good ($N_{\rm obs} \ge
5$, separation $\lesssim$ 6~arcmin, convergent cluster distance/group
redshift, distance class $<$ 6), such as the match between LCLG-42~010
and Abell~2758, and that others are not so good (separation $\gtrsim$
6~arcmin, widely divergent cluster distance/group redshift). Of the
latter, some of the LCRS groups are likely in the foreground
(LCLG-39~172, LCLG-39~202) or in the background (LCLG-03~126,
LCLG-39~256, LCLG-42~052, LCLG-42~234).  Furthermore, some Abell
clusters have been split into two or more groups by the
``friends-of-friends'' algorithm (Abell~1200, S418, 2969, S281, S253,
S286).  All in all, however, the matchups are not too bad, especially
for those groups with $N_{\rm obs} > 3$ observed members and those
clusters of Abell distance class $D < 6$.

If we consider just the non-split $N_{\rm obs} > 3, D < 6$ matches, we find by
least-squares fit that 
\begin{eqnarray}
C_{\rm grp} & \sim & 0.19C_{\rm ACO} + 12 \nonumber
\end{eqnarray}
(eq. [41]; see Fig.~20).  Such a poor correspondence between $C_{\rm
grp}$ and $C_{\rm ACO}$ may be disheartening at first --- at least
until one realizes that independent measures of $C_{\rm ACO}$ among 
Abell, Corwin, and Olowin {\em themselves\/} often had random
and systematic offsets of up to 50 galaxy counts (see ACO, Figs.~6 \&
7).  Furthermore, ACO stress that the measured counts for individual
clusters are nearly meaningless.  Therefore, we also stress that
$C_{\rm grp}$ (or the corresponding $C_{\rm ACO}$ from eq. [31]) for
individual LCRS groups will likely be very noisy; it is better to
consider mean or median values of these estimates for sets of LCRS
groups.

For instance, the median $C_{\rm grp}$ for LCRS groups is 10.8
(Fig.~19).  This value for $C_{\rm grp}$ indicates a median $C_{\rm
ACO} \sim -6$, which implies that LCRS groups, on average, can be
thought of as very poor clusters.  (The negative values for the median
counts is based in part to the different means of background
subtraction used here and in ACO.)  Clearly, the LCRS groups do define
an environment intermediate between that of isolated galaxies and that
of rich clusters.

\section{Summary and Conclusions}

We have presented in this paper a catalogue of loose groups within the
LCRS.  These groups were extracted from the LCRS galaxy catalogue by
means of a standard Huchra-Geller ``friends-of-friends'' percolation
algorithm, modified for comoving distances and for the field-to-field
sampling variations characteristic of this redshift survey.  

Internal comparisons of characteristics within the LCRS group
catalogue indicated some minor differences between groups extracted
from the 50-fiber fields and those extracted from the 112-fiber
fields.  We attributed these differences to some small but
still-hidden residual selection bias between galaxies in the 50-fiber
and in the 112-fiber fields.  Since groups in the 50-fiber sample
comprise only $\approx 20\%$ of the LCRS group catalogue, and since
the properties of both the 50-fiber and the 112-fiber groups fall
within the general regime of other group catalogues, we found this
discrepancy to be of only minor importance.

External comparison of the LCRS group catalogue with nine other group
catalogues, all based upon other redshift surveys, showed that the
general properties of LCRS groups are quite typical of current group
catalogues.  Nonetheless, it, along with the ESP group catalogue
(RZZ99), is the only group catalogue based upon a redshift survey
covering a reasonably fair sample of the local Universe.  Therefore,
the properties of the LCRS group catalogue, containing groups from a
wide range of environments, should be among the least biased to date.

Matchups of the LCRS groups with Abell clusters indicated, not
surprisingly, that, on average, these groups are much poorer than
Abell-class clusters, and therefore that LCRS groups do indeed inhabit
a range of parameter space intermediate to that of individual galaxies
and to that of rich clusters.

We therefore conclude that this catalogue will be useful for a variety
of studies requiring an unbiased census of loose groups, including the
measurement of the luminosity function of group members versus that of
field galaxies, the investigation of various morphology-environment
relations, and the study of the clustering of groups.

\acknowledgments

We thank the anonymous referee for the many useful comments.

DLT would also like to thank Marcio Maia, Ben Moore, Richard Nolthenius, 
Massimo Ramella, and Roberto Trasarti-Battistoni for valuable discussions
--- by e-mail or in person --- concerning their group catalogues.

The Las Campanas Redshift Survey has been supported by NSF grants AST
87-17207, AST 89-21326, and AST 92-20460.  This research has made use
of the NASA/IPAC Extragalactic Database (NED), which is operated by
the Jet Propulsion Laboratory, Caltech, under contract with the
National Aeronautics and Space Administration.

This work was supported in part by the US Department of Energy under
contract No. DE-AC02-76CH03000.

\newpage

\appendix

\section{Calculating $n^{\rm exp}(f,D)$ and $n^{\rm exp}_{\rm fid}$ for the LCRS}

We estimate $n^{\rm exp}(f,D)$ via 
\begin{equation}
n^{\rm exp}(f,D) = F \times \phi^* 
\int_{L_{\rm min}}^{L_{\rm max}} 
\left( \frac{L}{L^*} \right)^{\alpha} e^{-L/L^*} d\left( \frac{L}{L^*} \right) . 
\end{equation}
Equation~A1 is just the Schechter (1976) luminosity function
multiplied by a corrective factor $F$ and integrated over the interval
$L_{\rm min} \leq L \leq L_{\rm max}$.  $L_{\rm min}$ and $L_{\rm
max}$ are the extremal luminosities observable at a comoving distance
$D$ under the flux and luminosity limits imposed on field $f$, $F$ is
the field-to-field sampling fraction for field $f$, and $\phi^*$,
$L^*$, and $\alpha$ are the standard Schechter function parameters.

The value for $n^{\rm exp}_{\rm fid}$ is calculated in the same manner, 
\begin{equation}
n^{\rm exp}_{\rm fid} = F \times \phi^* 
\int_{L_{\rm min}}^{L_{\rm max}}
\left( \frac{L}{L^*} \right)^{\alpha} e^{-L/L^*} d\left( \frac{L}{L^*} \right), 
\end{equation}
where the values for $F$, $L_{\rm min}$, $L_{\rm max}$, $\phi^*$,
$L^*$, and $\alpha$ are those for the fiducial field.

As in Lin et al.\ (1996) and Tucker et al.\ (1997), we make use of two
different LCRS luminosity functions.  The first version is the standard
LCRS luminosity function; it best describes the data from the 112-fiber
fields and the Northern Galactic Cap 50-fiber fields.  Since these data
compose $\sim$ 90\% of the full LCRS sample, it is this version of the
luminosity function which we use for our fiducial field. The Schechter 
parameters for this luminosity function are as follows: 
\begin{equation}
\alpha = -0.70, M^* = -20.29 + 5 \log h, \phi^* = 0.019h^3~\mbox{Mpc}^{-3}
\end{equation}
($M^*$ is the absolute magnitude equivalent of $L^*$).
 
The measured luminosity function for the 50-fiber Southern Galactic
Cap data differs significantly from that of the other LCRS data (Lin
et al.\ 1996).  The reason behind this difference has never been fully
resolved, but it is thought to be the effect of subtle selection
problems in the very early LCRS data.  We use the following values
to describe the luminosity function of these Southern 50-fiber data:
\begin{equation}
\alpha = -0.74, M^* = -20.55 + 5 \log h, \phi^* = 0.016h^3~\mbox{Mpc}^{-3} .
\end{equation} 

[Note: Both Equations~A1 and A2 are actually simplified forms of the
integral found in the group-finding code.  The integral in the code
also includes effects due to apparent magnitude and surface brightness
incompleteness and due to central surface brightness selection;
furthermore, in the code, this integral is convolved with a gaussian
flux error of $\sigma = 0.1$~mag.  For a detailed discussion of these
additional selection effects, see \S~3.2 of Lin et al.\ 1996.]

\section{Calculating $L_{\rm tot}$ and $\sigma_{L_{\rm tot}}$ for Las Campanas Loose Groups}

To correct for selection effects, $L_{\rm tot}$ is calculated by means
of the following equation:
\begin{equation}
L_{\rm tot} = \sum_{i=1}^{N_{\rm obs}} {\mathcal{L}}_i , 
\end{equation}
where
\begin{equation}
{\mathcal{L}}_i \equiv \left( \frac{n^{\rm tot}_{\rm lum}}{n^{\rm exp}_{\rm lum}(f_i,D(z_i))} \right) \times L_i ,
\end{equation}
where $N_{\rm obs}$ is the number of {\it observed\/} members in the
group, $L_i$ is the luminosity of group member $i$, $n^{\rm tot}_{\rm
lum}$ is the {\it total\/} expected luminosity density of {\it all\/}
galaxies in the local Universe, and $n^{\rm exp}_{\rm
lum}(f_i,D(z_i))$ is the expected luminosity density for only those
galaxies that would lie within the photometric boundaries of LCRS
field $f_i$ at a comoving distance $D(z_i)$.

To estimate $n^{\rm tot}_{\rm lum}$, we integrate the
luminosity-weighted Schechter function over all luminosities ($0 \le L
\le \infty$):
\begin{eqnarray}
n^{\rm tot}_{\rm lum} & = & \phi^* L^* \int_{0}^{\infty} \left(
\frac{L}{L^*} \right)^{\alpha+1} e^{-L/L^*} d\left( \frac{L}{L^*} \right) \\
                      & = & \phi^* L^* \Gamma(\alpha+2), \nonumber
\end{eqnarray}
where $\phi^*$, $\alpha$, and $L^*$ are the standard Schechter
parameters, and where $\Gamma$ is the complete gamma function from
mathematics (Abramowitz \& Stegun 1970).  

We estimate $n^{\rm exp}_{\rm lum}(f_i,D(z_i))$ via
\begin{equation}
n^{\rm exp}_{\rm lum}(f_i,D(z_i)) = F(f_i) \times \phi^* L^* \int_{L_{\rm
min}}^{L_{\rm max}} \left( \frac{L}{L^*} \right)^{\alpha+1} e^{-L/L^*}
d\left( \frac{L}{L^*} \right) . 
\end{equation}
Similar to the cases of equations~A1 and A2, equation~B4 is just the
luminosity-weighted Schechter function multiplied by a corrective
factor $F(f_i)$ and integrated over the interval $L_{\rm min} \leq L
\leq L_{\rm max}$.  $L_{\rm min}$ and $L_{\rm max}$ are the extremal
luminosities observable at redshift $z_i$ under the given flux and
luminosity limits imposed on field $f_i$; $F(f_i)$ is the
field-to-field sampling fraction for field $f_i$.  [Note: Equation~B4
is actually a simplified form of the integral found in the
group-finding code.  The integral in the code also includes effects
due to apparent magnitude and surface brightness incompleteness and
due to central surface brightness selection; furthermore, in the code,
this integral is convolved with a gaussian flux error of $\sigma =
0.1$~mag.  For a detailed discussion of these additional selection
effects, see \S~3.2 of Lin et al.\ 1996.]

The rms error in $L_{\rm tot}$ is estimated by summing the individual
contributions from the rms errors in ${\mathcal{L}}_i$ in quadrature,
yielding
\begin{equation}
\sigma_{L_{\rm tot}} = N_{\rm obs} 
\sqrt{
\frac{<{\mathcal{L}}_i^2> - <{\mathcal{L}}_i>^2}{N_{\rm obs} - 1} 
} , 
\end{equation}
where
\begin{equation}
<{\mathcal{L}}_i> = 
\frac{\sum_{i=1}^{N_{\rm obs}} {\mathcal{L}}_i}{N_{\rm obs}}
\end{equation}
and
\begin{equation}
<{\mathcal{L}}_i^2> = 
\frac{\sum_{i=1}^{N_{\rm obs}} {\mathcal{L}}_i^2}{N_{\rm obs}} .
\end{equation}

\section{Estimating Abell Counts for Las Campanas Loose Groups}

We wish to make a quantitative, unbiased estimate of the richnesses of
the Las Campanas loose groups. A useful and historically motivated
method is to calculate their Abell counts, $C$.  $C$ was defined by
Abell (1958) to be the number of galaxies, corrected for background
contamination, in the magnitude interval $m_3$ to $m_3 + 2$ that lie
within a $1.7~{\rm arcmin}/z$ ($\approx 1.5~h^{-1}$~Mpc) projected
separation of a cluster's center; the magnitude $m_3$ is the magnitude
of the third brightest cluster member.  Due to the sampling
characteristics and the relatively small range of apparent magnitudes
within the LCRS, we cannot use Abell's definition directly.  We must
take a more circuitous path, via the use of the Schechter function for
groups and clusters of galaxies.

For simplicity, consider a group which lies entirely within a single
field.  In this case, $n^{\rm obs}_{1.5}$, the observed number of
galaxies within a $1.5~h^{-1}$~Mpc projected separation of a group's
barycenter, should fit the relation
\begin{equation}
n^{\rm obs}_{1.5} = F(f_i) \times n^*_{1.5}  \int_{L_{\rm min}}^{L_{\rm max}} 
\left(\frac{L}{L^*} \right)^{\alpha} e^{-L/L^*} d\left( \frac{L}{L^*} \right),
\end{equation}
which is just the integral of the Schechter function over the
luminosity range $L_{\rm min} \leq L \leq L_{\rm max}$.  As in
Appendix~B, $L_{\rm min}$ and $L_{\rm max}$ are the extremal
luminosities observable at redshift $z_i$ under the given flux and
luminosity limits imposed on field $f_i$, and $F(f_i)$ is the
field-to-field sampling fraction for field $f_i$.  Here, $n^*_{1.5}$
is a normalization factor for galaxy number counts within a projected
radius of $1.5~h^{-1}$~Mpc of a given group's barycenter; it is a
counterpart to $\phi^*$, which is used for the field galaxy luminosity
function.  The value for $n^*_{1.5}$ is itself a measure of richness;
the richer the group or cluster, the higher the value of $n^*_{1.5}$
(assuming constant $\alpha$ and $L^*$).  We will use $n^*_{1.5}$ to
make an estimate of group Abell richness.  Observationally, we can
calculate a value for $n^{\rm obs}_{1.5}$ in equation~C1 by
subtracting off the estimated number of interlopers ($n_{\rm I}$;
equation~14) from the total number of observed galaxies in a group
within $1.5~h^{-1}$~Mpc of the group's barycenter ($N_{\rm
obs}^{1.5}$),
\begin{equation}
n^{\rm obs}_{1.5} = \sum_{1}^{N_{\rm obs}^{1.5}} [ 1 - n_{\rm I}(f_i,cz_i) ]
\end{equation}
Then, placing this result into equation~C1, 
\begin{equation}
n^*_{1.5} = \frac{n^{\rm obs}_{1.5}} {F(f_i) \int_{L_{\rm min}}^{L_{\rm max}} 
\left( \frac{L}{L^*} \right)^{\alpha} e^{-L/L^*} d\left( \frac{L}{L^*} \right)}.
\end{equation}
Since the LCRS is a $\sim$ 75\%-sampled redshift catalogue with both
bright and faint apparent magnitude cutoffs, it is quite possible that
the third brightest group member may not be in the LCRS spectroscopic
sample.  Therefore, an estimate for the absolute magnitude of the
third brightest group member must be computed.  We can do this by
integrating the group's luminosity function,
\begin{eqnarray}
n_{\rm e}(\ge L) & = & n^*_{1.5} \int_{L}^{\infty} 
                       \left (\frac{L}{L^*} \right)^{\alpha} e^{-L/L^*} 
                       d\left( \frac{L}{L^*} \right) \\
                      & = &  n^*_{1.5} \Gamma(\alpha + 1,L/L^*), \nonumber
\end{eqnarray}
where $\Gamma(\alpha+1,L/L^*)$ is the incomplete gamma function
(Abramowitz \& Stegun 1970), and solving the equation
\begin{equation}
n_{\rm e}(\geq L(m_3)) = 3 = n^*_{1.5} \Gamma(\alpha + 1,L(m_3)/L^*)
\end{equation}
numerically for the luminosity of the third brightest cluster member,
$L(m_3)$.  Finally, taking our estimate of $n^*_{1.5}$ from
equation~C3 and our estimate of $L(m_3)$ from equation~C5, we can
calculate the group's Abell Richness, which we will call $C_{\rm
grp}$,
\begin{equation}
C_{\rm grp} = n^*_{1.5}  \int_{L(m_3)}^{L(m_3 + 2)} \left(
\frac{L}{L^*} \right)^{\alpha} e^{-L/L^*} d\left( \frac{L}{L^*} \right), 
\end{equation}
where $L(m_3 + 2)$ is the luminosity associated with the apparent
magnitude $m_3 + 2$ at the redshift of the group in question.

A further complication exists in that the linking radius employed in
the ``friends-of-friends'' percolation algorithm is only $\sim$
1~$h^{-1}$~Mpc at the fiducial redshift $z_{\rm fid}$; for some tightly
configured groups containing only a few observed members, the full
Abell radius of 1.5~$h^{-1}$~Mpc (projected) may not be completely
searched, resulting in underestimates of the Abell counts $C$.  Since
most LCRS groups are not very tightly configured (consider the mean
pairwise separations, $R_{\rm p}$), this is not likely to be a significant
effect for the catalogue as a whole.  In a similar case, APM clusters,
which number counts only within a projected radius of 0.75~$h^{-1}$~Mpc
from the cluster center, have been shown to underestimate their Abell
counts typically by only $\sim$ 20\% or less (Bahcall \& West
1992).  The LCRS percolation algorithm, with its relatively large
linking parameter $D_{\rm L}$ (typically $\gtrsim 1h^{-1}$~Mpc), should
perform much better than do the APM counts. 

When compared with actual Abell clusters within the LCRS volume, we
find the following relation between the LCRS group counts estimate,
$C_{\rm grp}$, and the revised Abell cluster values given by ACO,
$C_{\rm ACO}$:
\begin{eqnarray}
C_{\rm grp} & \sim & 0.19C_{\rm ACO} + 12 \nonumber
\end{eqnarray}
(eq. [41]).  We discuss this relation in more detail in \S~6.

\clearpage


\begin{figure}
\caption{The LCRS survey pattern for the Northern (top) and the
Southern (bottom) Galactic Cap regions.  Lightly shaded regions denote
fields observed with the 50-fiber MOS and darkly shaded regions fields
observed with the 112-fiber MOS.  Declination and right ascension
coordinates are equinox 1950.0}
\end{figure}

\clearpage

\begin{figure}
\plotfiddle{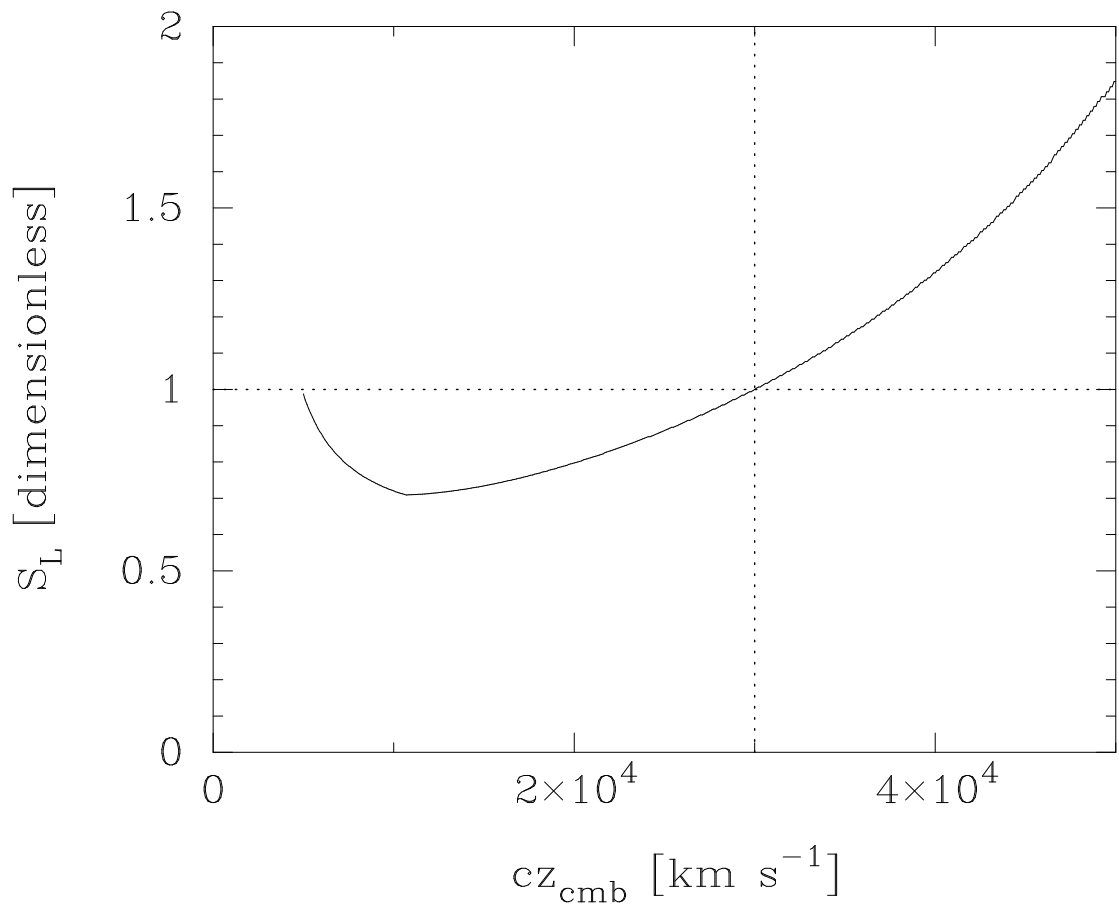}{5.00in}{-90}{75}{75}{-295}{500}
\caption{Variation of the linking scale $S_{\rm L}$ with velocity for
the fiducial field ($15.0 \leq R < 17.7$, 100\% sampling), assuming
$\alpha = -0.70$, $M^* = -20.29 + 5 \log h$, and $\phi^* =
0.019h^3$~Mpc$^{-3}$.  The dotted lines indicate the locus of $S_{\rm
L} = 1$ and $cz_{\rm cmb} = 30,000$~km~s$^{-1}$ (the fiducial
velocity).}
\end{figure}

\clearpage

\begin{figure}
\plotfiddle{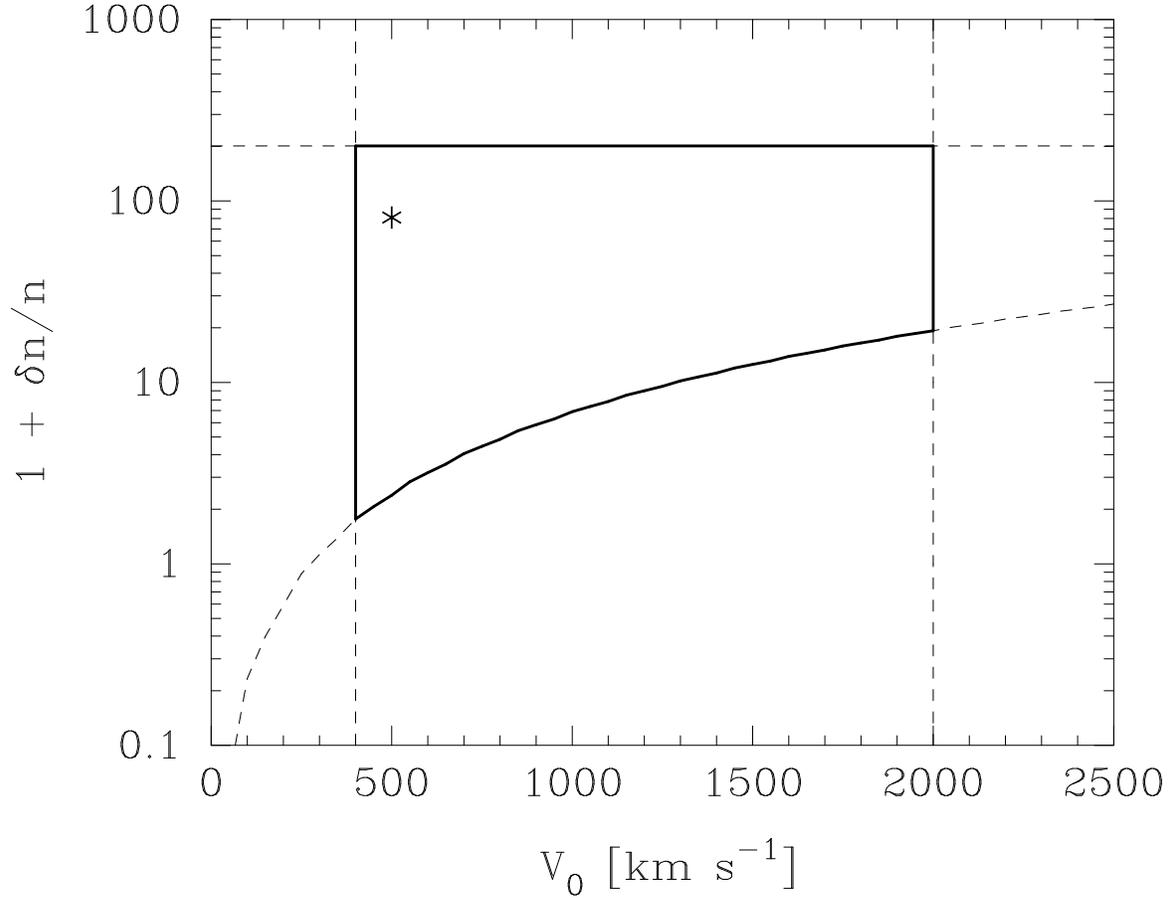}{5.00in}{-90}{75}{75}{-295}{500}
\caption{Group selection parameters.  The region of reasonable search
parameter values is bound in solid (see text for details).  The
asterisk indicates the final choice used in extracting the LCRS group
catalogue: $\delta n/n = 80$ ($\Longleftrightarrow D_0 =
0.715~h^{-1}$~Mpc) and $V_0 = 500$~km~s$^{-1}$.}
\end{figure}

\clearpage

\begin{figure}
\caption{The number of interlopers per galaxy, $n_{\rm I}$, at the
redshift of each galaxy in the LCRS (assuming the final values for
$\delta n/n$ ($D_0$) and $V_0$).  The median $n_{\rm I}$ is 0.17
interlopers per galaxy.}
\end{figure}

\clearpage

\begin{figure}
\plotfiddle{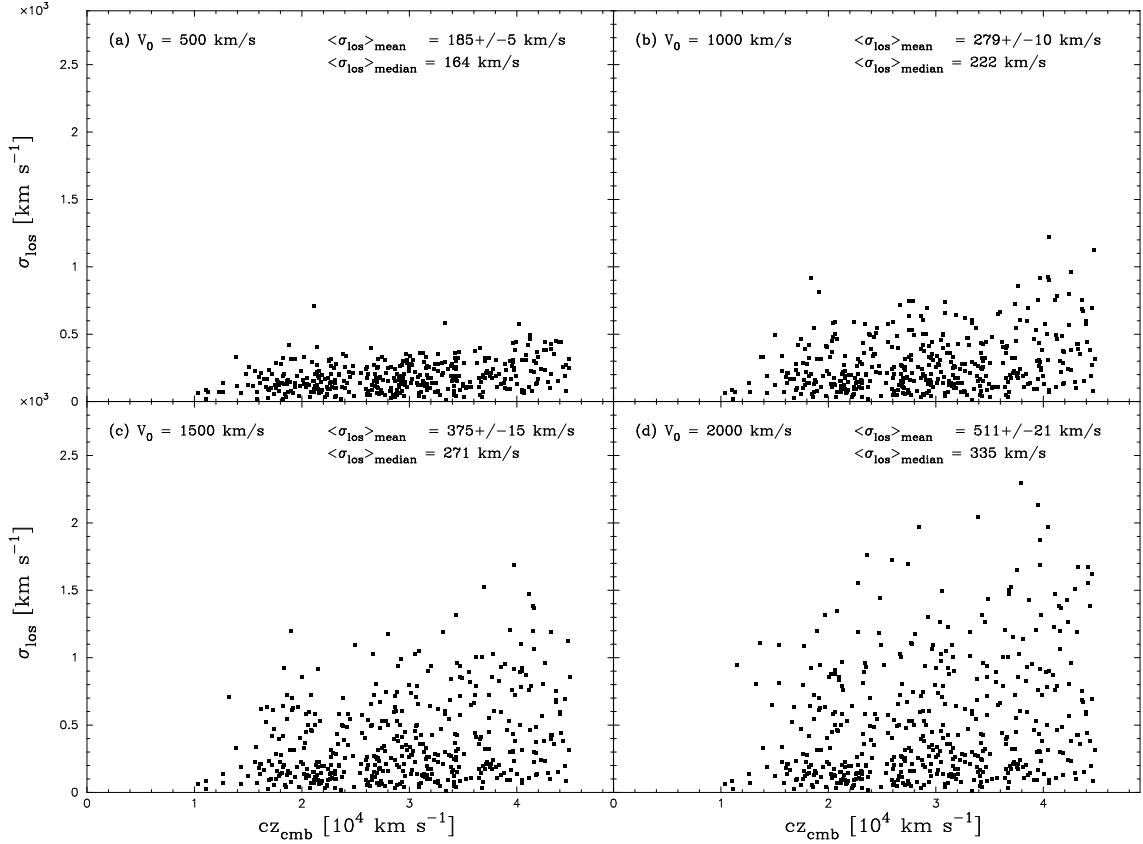}{5.00in}{-90}{75}{75}{-290}{475}
\caption{Group line-of-sight velocity dispersions vs.\ redshift for
$\delta n/n = 80$ and
(a) $V_0 = 500$~km~s$^{-1}$
(b) $V_0 = 1000$~km~s$^{-1}$,
(c) $V_0 = 1500$~km~s$^{-1}$, and
(d) $V_0 = 2000$~km~s$^{-1}$.
(N.B.: Only groups meeting the requirements of a clean sample
--- i.e., groups with $\sigma_{\rm los} > 0$~km~s$^{-1}$, with
barycenters $> 2R_{\rm p}$ from a slice edge, with crossing times $<$ a
Hubble time, and with no galaxies with a mock redshift --- were
included in these plots; for a more in-depth discussion of the
requirements of a clean sample, see \S~4.)}
\end{figure}

\clearpage

\begin{figure}
\caption{The distribution of galaxies in the LCRS Northern (top) and
Southern (bottom) Galactics out to $cz = 46,000$~km~s$^{-1}$ (to
include group members beyond the group catalogue $cz =
45,000$~km~s$^{-1}$ limit).  Only those galaxies having luminosity
$-22.5 \leq M_R - 5\log h < -17.5$ and lying within the LCRS official
geometric and photometric boundaries are plotted.  Red points are the
55-arcsec ``orphans,'' plotted with their mock velocities.
$N_{\rm tot}$ is the total number of galaxies plotted, 55-arcsec
``orphans'' included; $N_{\rm <55''}$ refers to the number of
55-arcsec ``orphans'' plotted.}
\end{figure}

\clearpage

\begin{figure}
\caption{Same as Figure~6, but only galaxies in $\delta n/n = 80$
groups are plotted.}
\end{figure}

\clearpage

\begin{figure}
\caption{The distribution of $\delta n/n = 80$ groups in the LCRS
Northern (top) and Southern (bottom) Galactic Caps.  Red symbols
indicate groups containing at least one 55-arcsec ``orphan''.  $N_{\rm
grp}$ refers to the total number of groups plotted, $N_{\rm <55''}$ to
the the number which contain 55-arcsec ``orphans''. (N.B.: The LCRS
group catalogue extends from $cz = 10,000$~km~s$^{-1}$ to $cz =
45,000$~km~s$^{-1}$; so the dearth of groups at $cz <
10,000$~km~s$^{-1}$ is not physical but merely the cutoff of the
catalogue.)}
\end{figure}

\clearpage

\begin{figure}
\plotfiddle{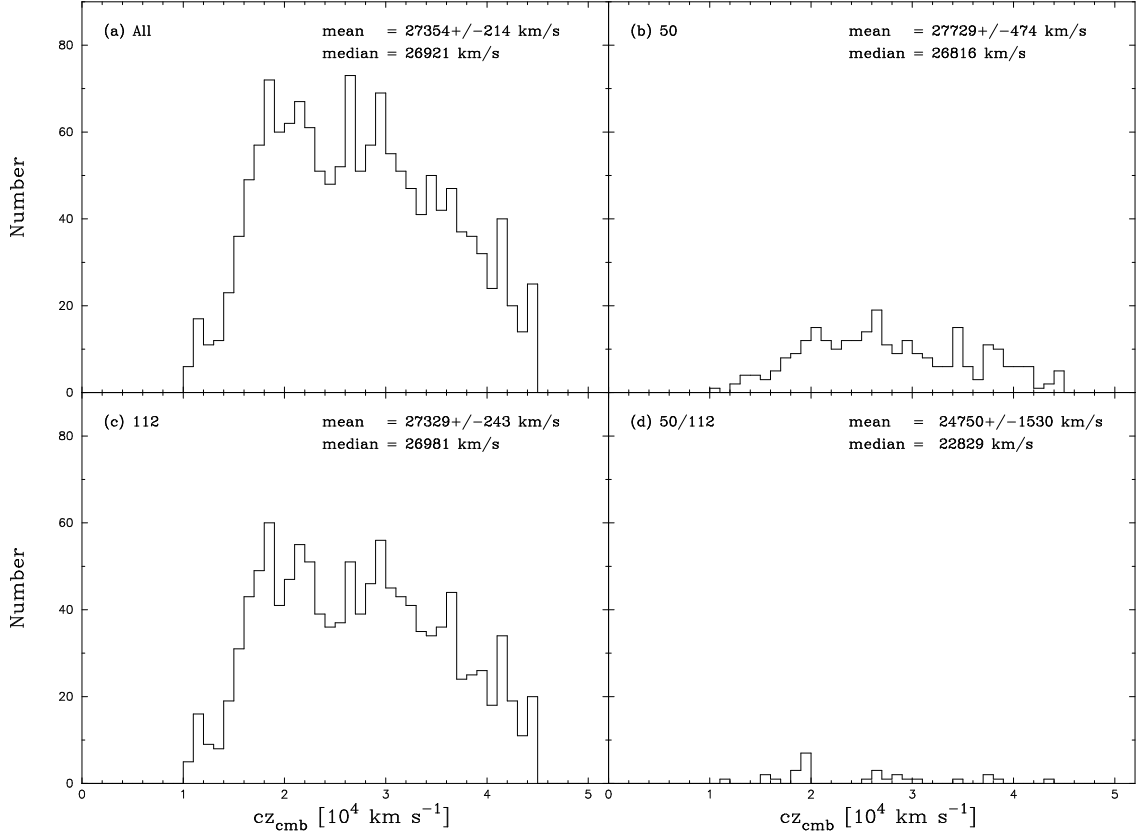}{5.00in}{-90}{75}{75}{-290}{475}
\caption{Distribution of group velocities from the full sample of 1495
groups: (a) all the groups, (b) just those groups in the 50-fiber
fields, (c) just those groups in the 112-fiber fields, (d) just those
groups which straddle a 50-/112-fiber field boundary.}
\end{figure}

\clearpage

\begin{figure}
\plotfiddle{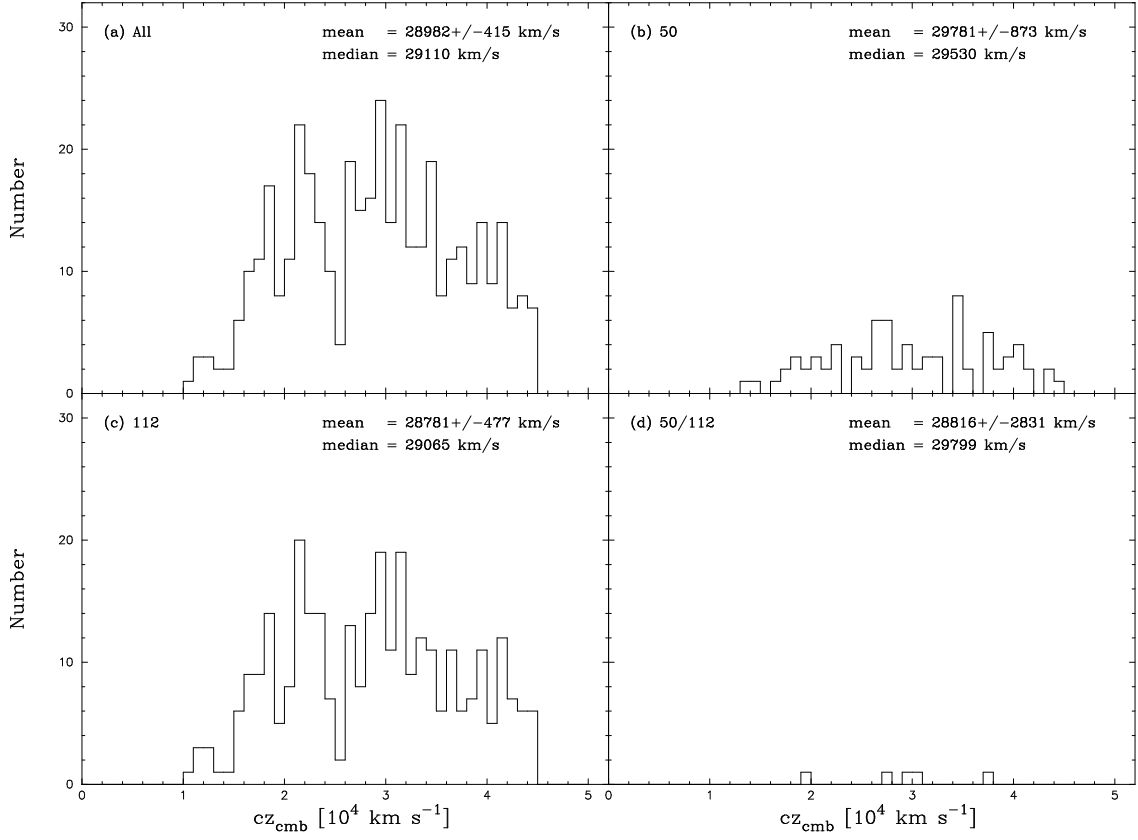}{5.00in}{-90}{75}{75}{-290}{475}
\caption{Distribution of group velocities from the clean sample
of 394 groups: (a) -- (d) are as in Figure~9.}
\end{figure}

\clearpage

\begin{figure}
\plotfiddle{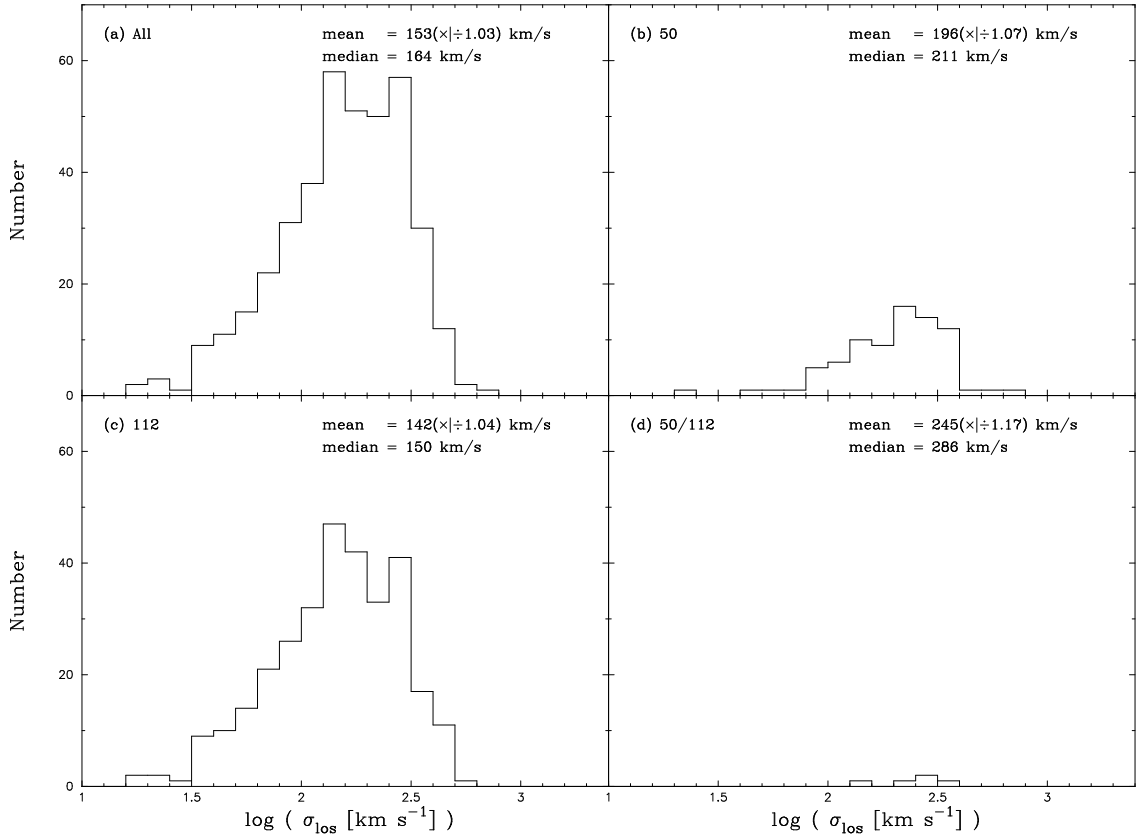}{5.00in}{-90}{75}{75}{-290}{475}
\caption{Distribution of line-of-sight velocity dispersions,
$\sigma_{\rm los}$, for LCRS groups in the clean sample: (a) --
(d) are as in Figure~9.}
\end{figure}

\clearpage

\begin{figure}
\plotfiddle{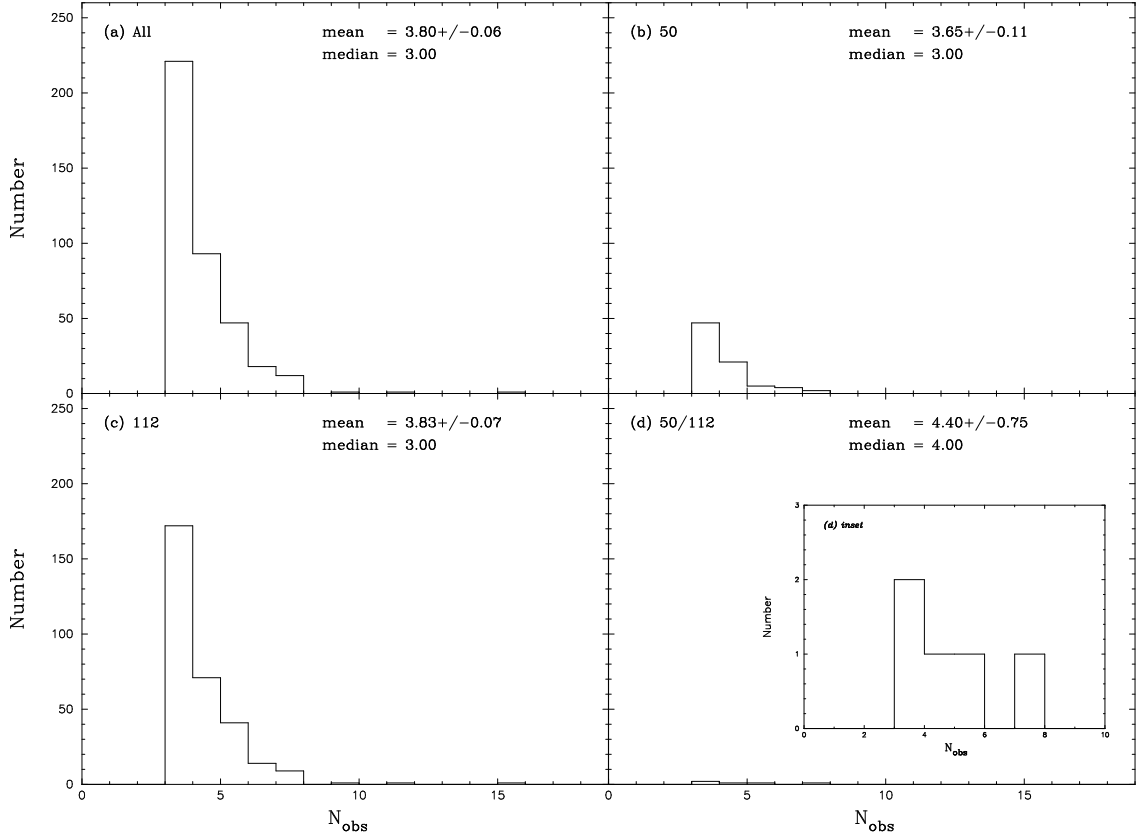}{5.00in}{-90}{75}{75}{-290}{475}
\caption{Distribution of the observed number of LCRS galaxies within a
group, $N_{\rm obs}$, for the LCRS groups in the clean sample:
(a) -- (d) are as in Figure~9; inset of (d) is a merely a blow-up to
aid the reader.}
\end{figure}

\clearpage

\begin{figure}
\plotfiddle{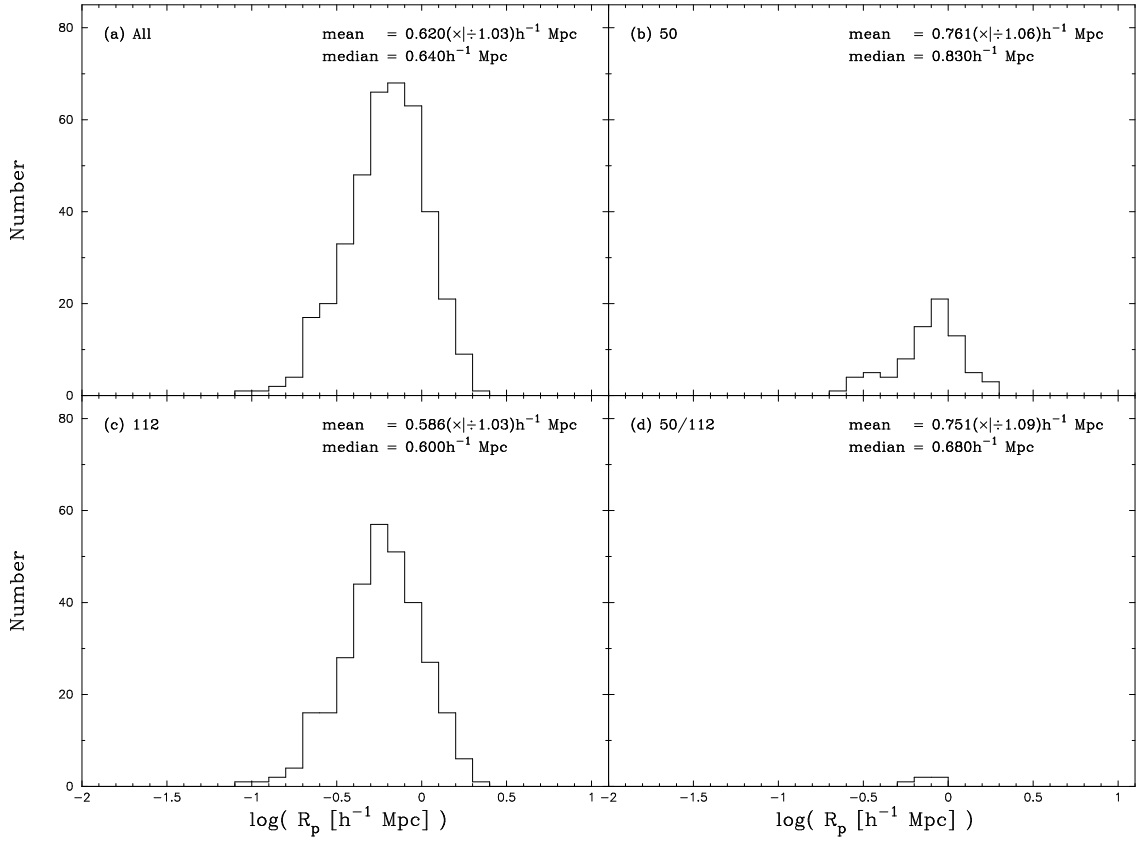}{5.00in}{-90}{75}{75}{-290}{475}
\caption{Distribution of mean pairwise separations, $R_{\rm p}$, for LCRS
groups in the clean sample: (a) -- (d) are as in Figure~9.}
\end{figure}

\clearpage

\begin{figure}
\plotfiddle{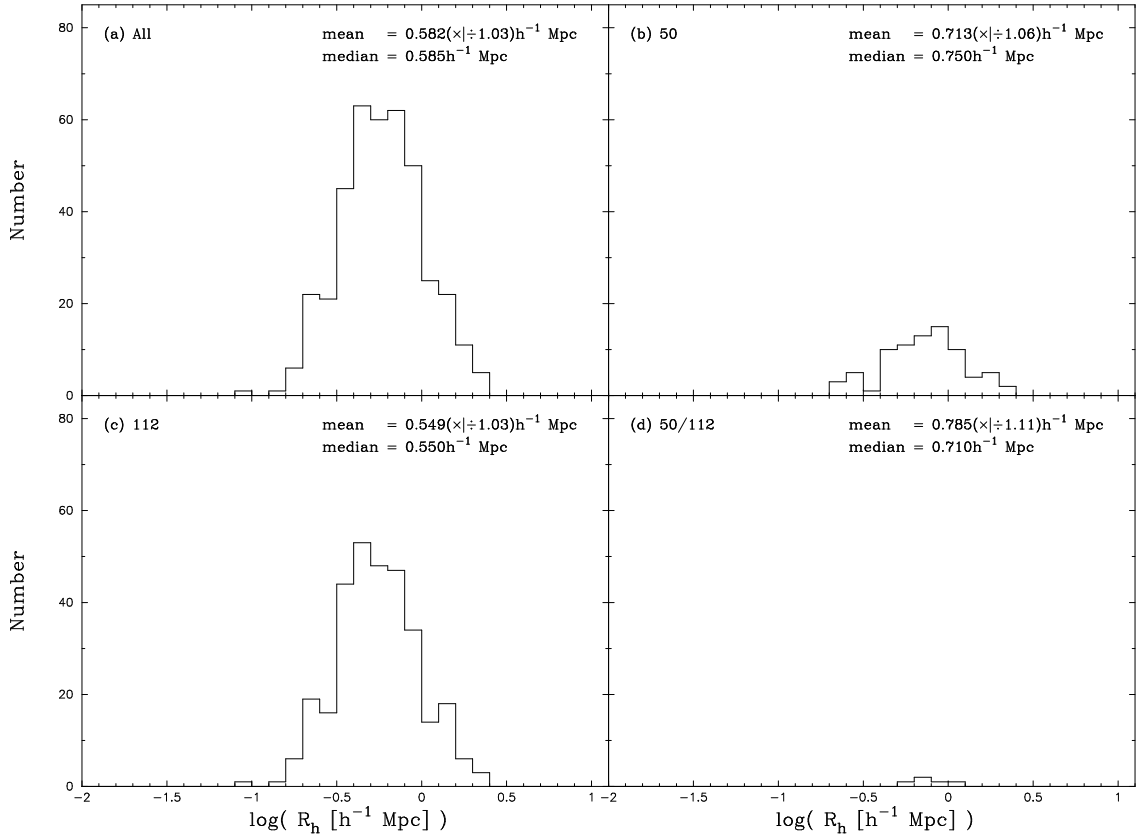}{5.00in}{-90}{75}{75}{-290}{475}
\caption{Distribution of harmonic radii, $R_{\rm h}$, for LCRS groups in
the clean sample: (a) -- (d) are as in Figure~9.}
\end{figure}

\clearpage

\begin{figure}
\plotfiddle{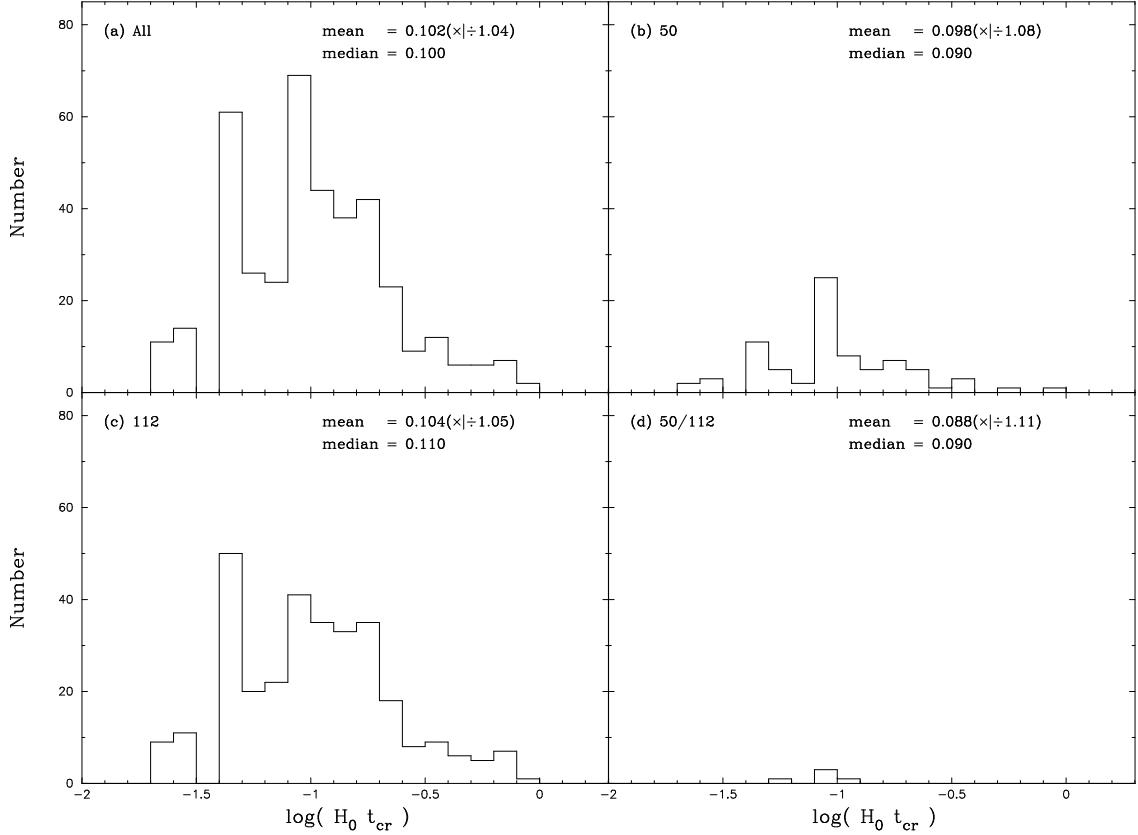}{5.00in}{-90}{75}{75}{-290}{475}
\caption{Distribution of virial crossing times, $t_{\rm cr}$, as a
fraction of the Hubble time ($H_0^{-1}$), for LCRS groups in the
clean sample: (a) -- (d) are as in Figure~9.  Following Gott \&
Turner (1977), groups with crossing times $t_{\rm cr} \lesssim 0.11
H_0^{-1}$ should have had enough time in the age of the Universe to
virialize completely [but cf.\ Diaferio et al.\ (1993)].}
\end{figure}

\clearpage

\begin{figure}
\plotfiddle{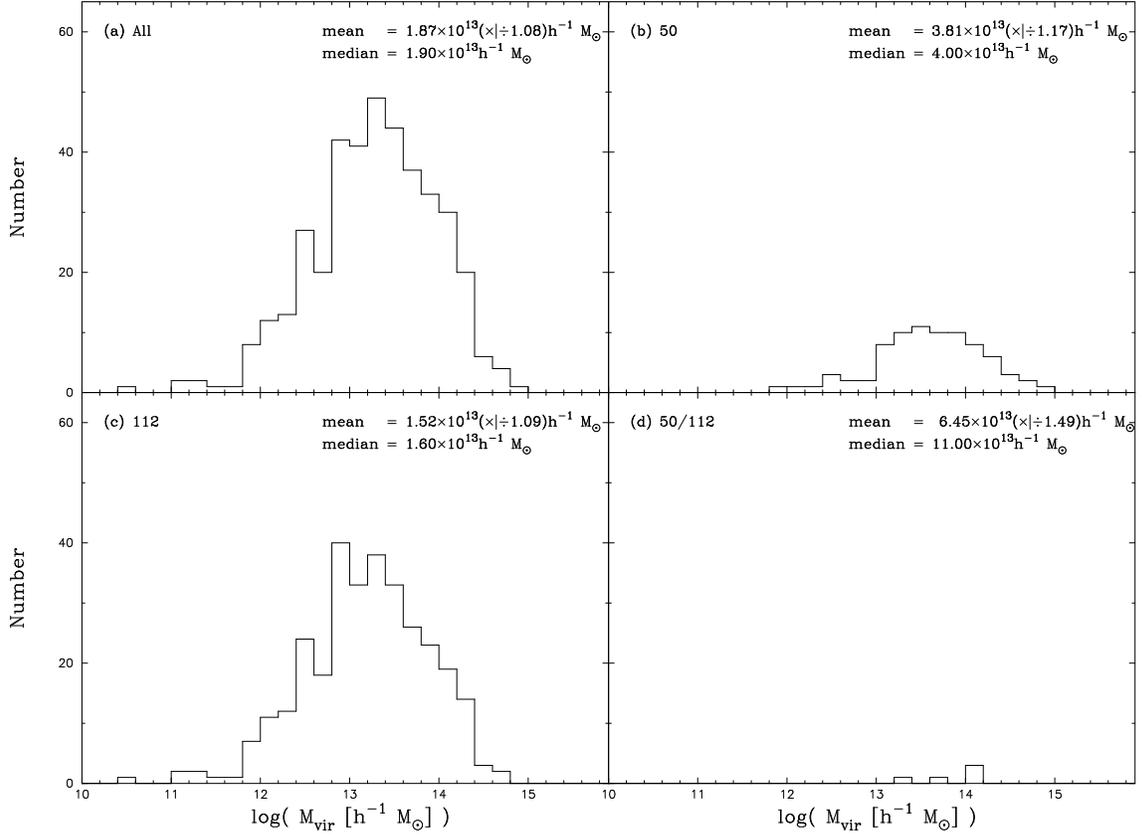}{5.00in}{-90}{75}{75}{-290}{475}
\caption{Distribution of virial masses, $M_{\rm vir}$, for LCRS groups
in the clean sample: (a) -- (d) are as in Figure~9.}
\end{figure}

\clearpage

\begin{figure}
\plotfiddle{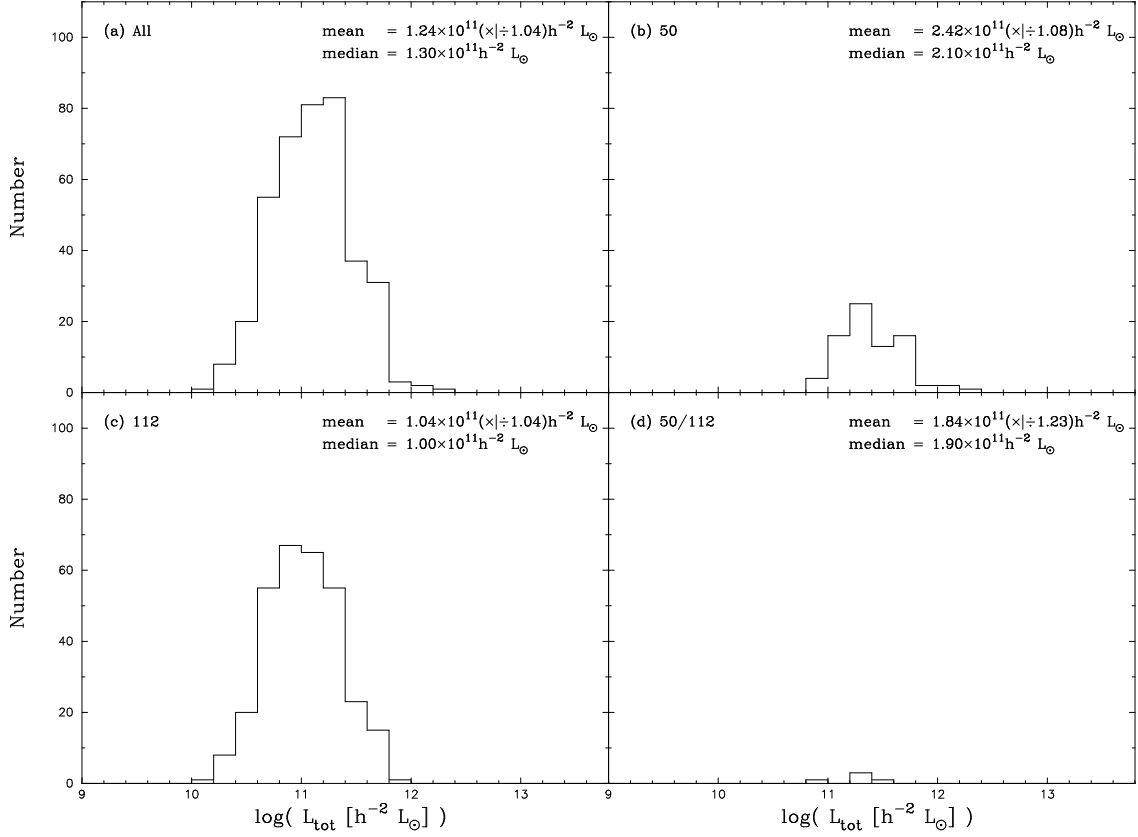}{5.00in}{-90}{75}{75}{-290}{475}
\caption{Distribution of estimated $R$-band total luminosities,
$L_{\rm tot}$, for LCRS groups in the clean sample: (a) -- (d)
are as in Figure~9.}
\end{figure}

\clearpage

\begin{figure}
\plotfiddle{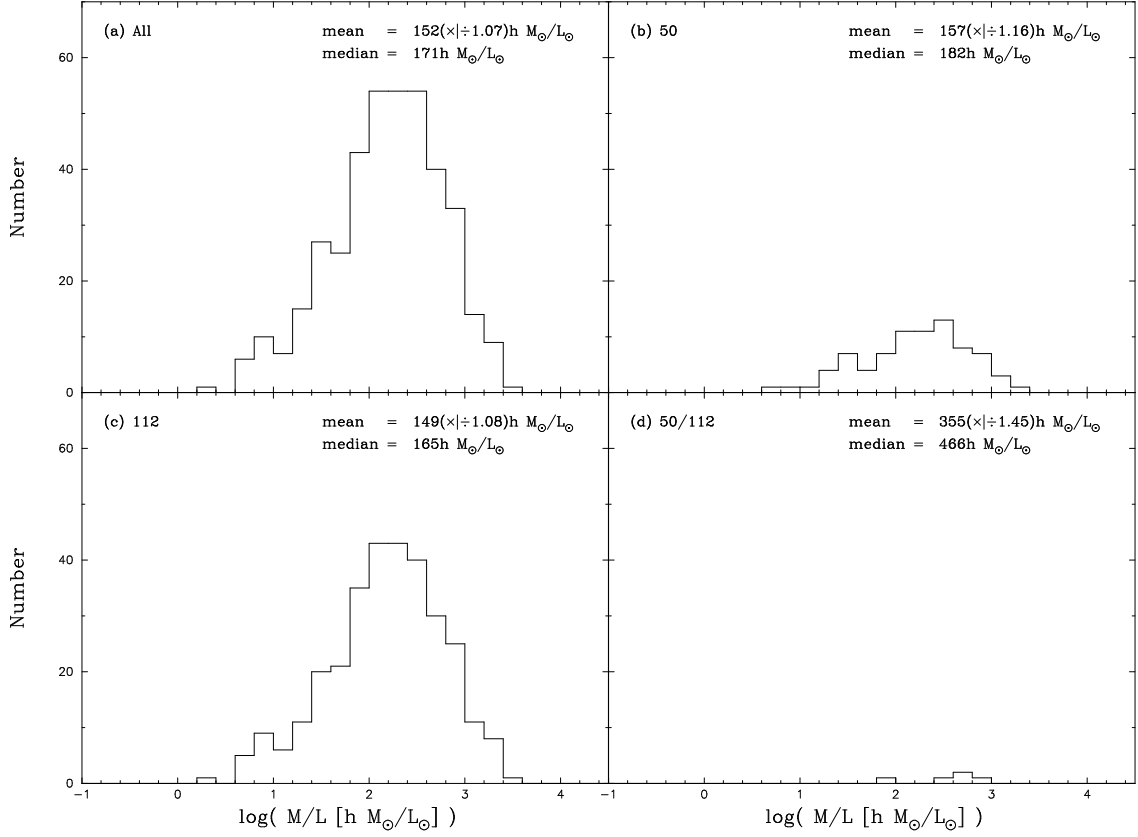}{5.00in}{-90}{75}{75}{-290}{475}
\caption{Distribution of estimated $R$-band mass-to-light ratios,
$M/L$, for LCRS groups in the clean sample: (a) -- (d) are as in
Figure~9.}
\end{figure}

\clearpage

\begin{figure}
\plotfiddle{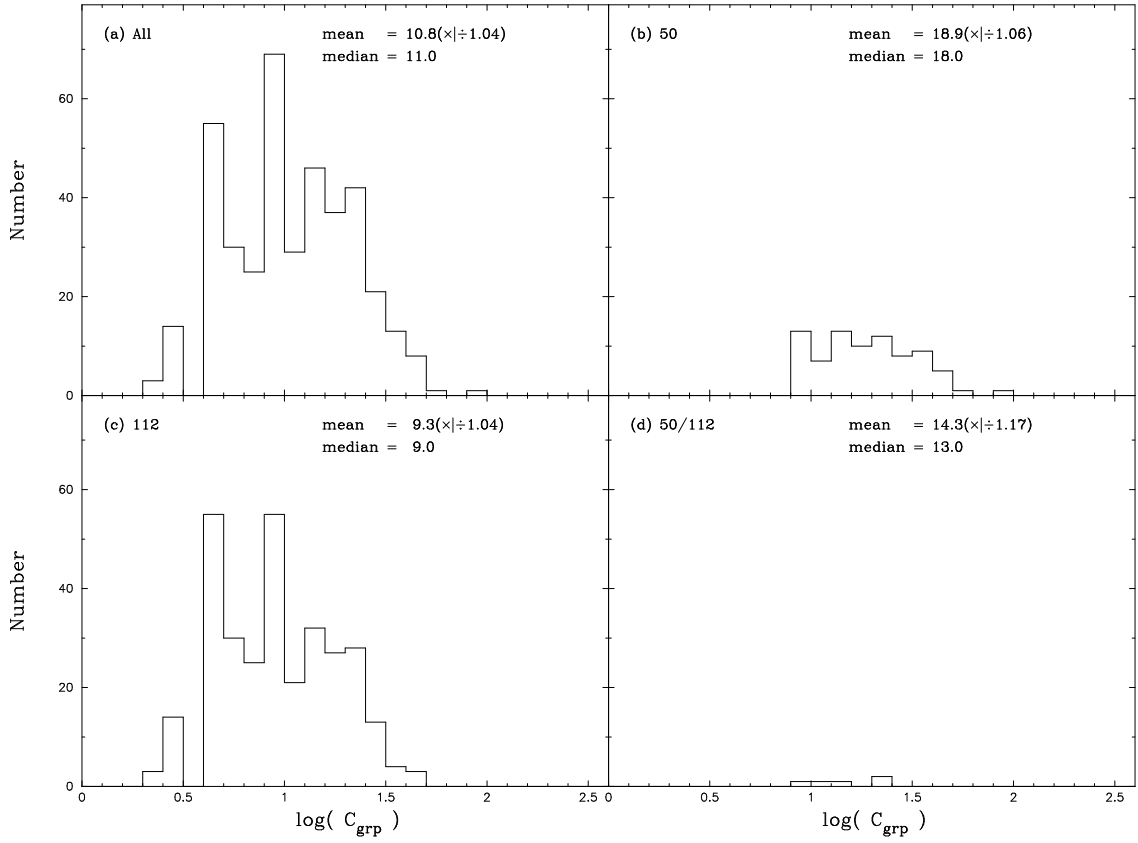}{5.00in}{-90}{75}{75}{-290}{475}
\caption{Distribution of Abell-like group counts (or richnesses),
$C_{\rm grp}$, for LCRS groups in the clean sample: (a) -- (d)
are as in Figure~9.}
\end{figure}

\clearpage

\begin{figure}
\plotfiddle{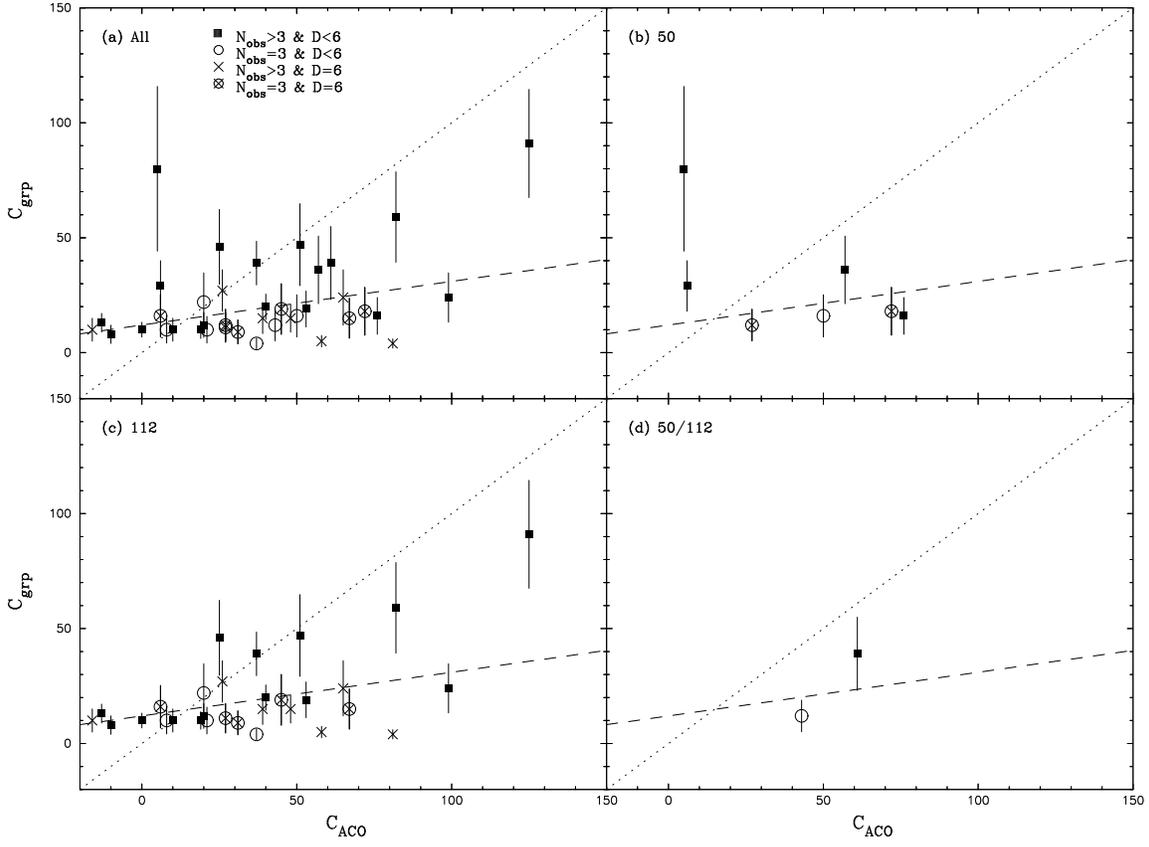}{5.00in}{-90}{75}{75}{-290}{475}
\caption{Group counts $C_{\rm grp}$ vs.\ Abell counts $C_{\rm ACO}$
for LCRS groups which are identified with an ACO cluster.  The sample
of LCRS groups used in the matchups was a superset of the clean
sample (groups including galaxies with mock velocities were also
included).  {\em Filled squares\/} denote groups with more than 3
observed members matched with ACO clusters of Abell distance class $D
< 6$; {\em open circles\/} denote groups with exactly 3 observed
members matched with distance class $D < 6$ ACO clusters; {\em $\times$'s\/}
denote groups with more than 3 observed members matched with distance
class $D=6$ ACO clusters; and {\em $\times$'ed circles\/} denote groups
having exactly 3 observed members identified with distance class $D=6$
ACO clusters.  The {\em dotted line\/} represents the locus of $C_{\rm
grp} = C_{\rm ACO}$; the {\em dashed line\/} represents the best-fit
line, $C_{\rm grp} = 0.19C_{\rm ACO} + 12$, for $N_{\rm obs} > 3, D <
6$ matches.  Plots (a) -- (d) are as in Figure~9.}
\end{figure}

\clearpage

\begin{figure}
\plotfiddle{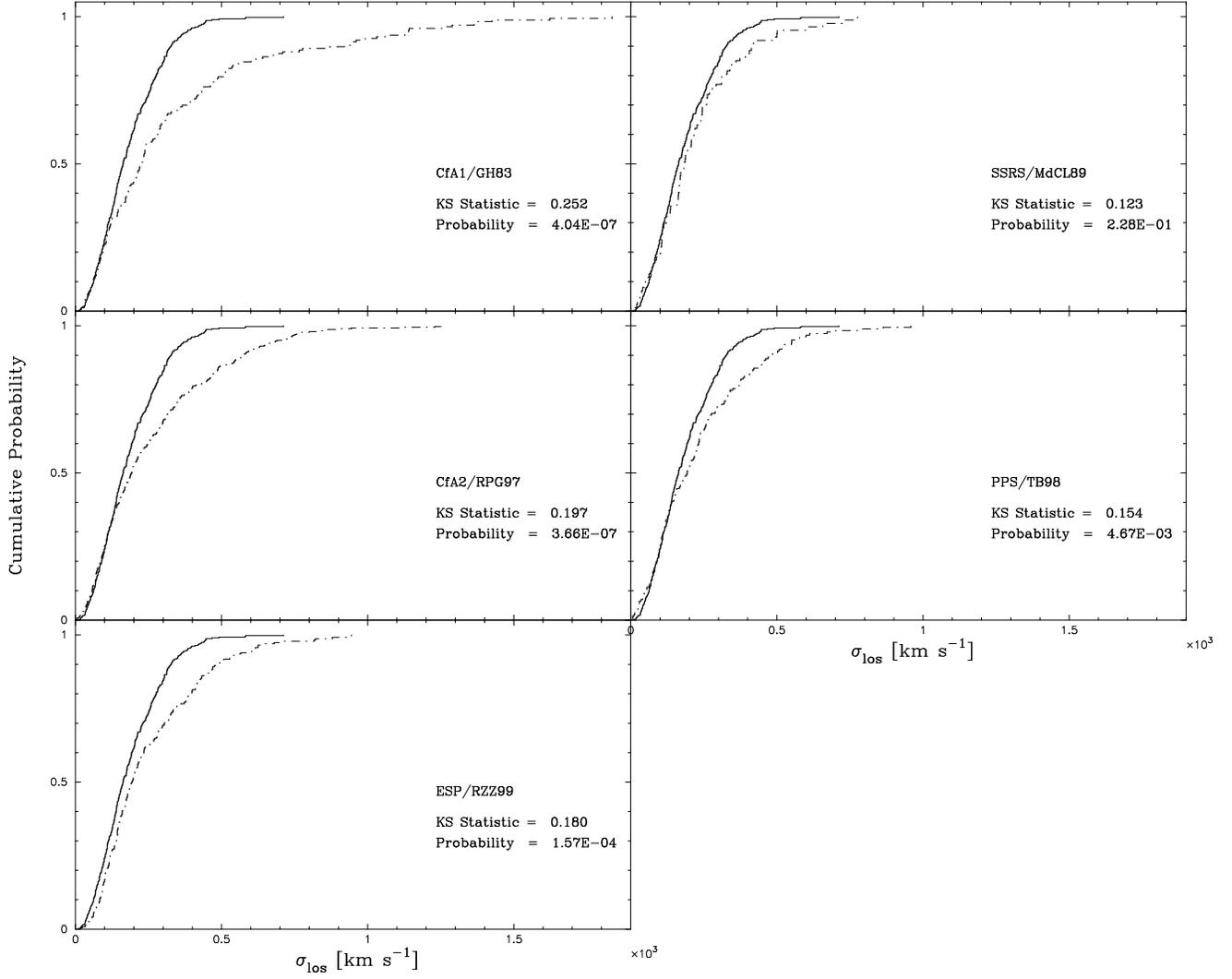}{5.00in}{-90}{75}{75}{-290}{475}
\caption{KS tests comparing the distribution of line-of-sight velocity
dispersions from various group catalogues ({\em dot-dashed line\/})
against that from the full LCRS group catalogue ({\em solid line\/}).}
\end{figure}

\clearpage

\begin{figure}
\plotfiddle{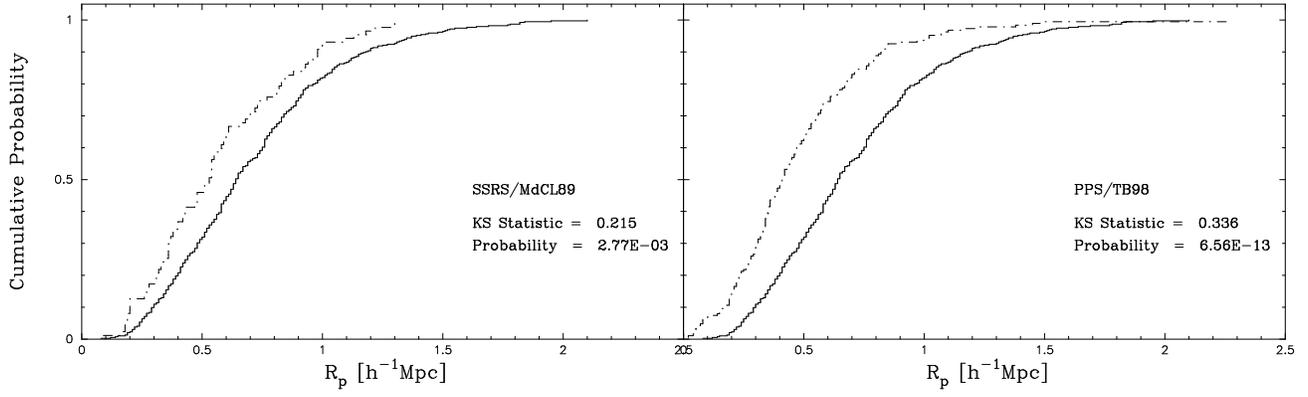}{5.00in}{-90}{75}{75}{-290}{475}
\caption{KS tests comparing the distribution of mean pairwise
separations from various group catalogues ({\em dot-dashed line\/})
against that from the full LCRS group catalogue ({\em solid line\/}).}
\end{figure}

\clearpage

\begin{figure}
\plotfiddle{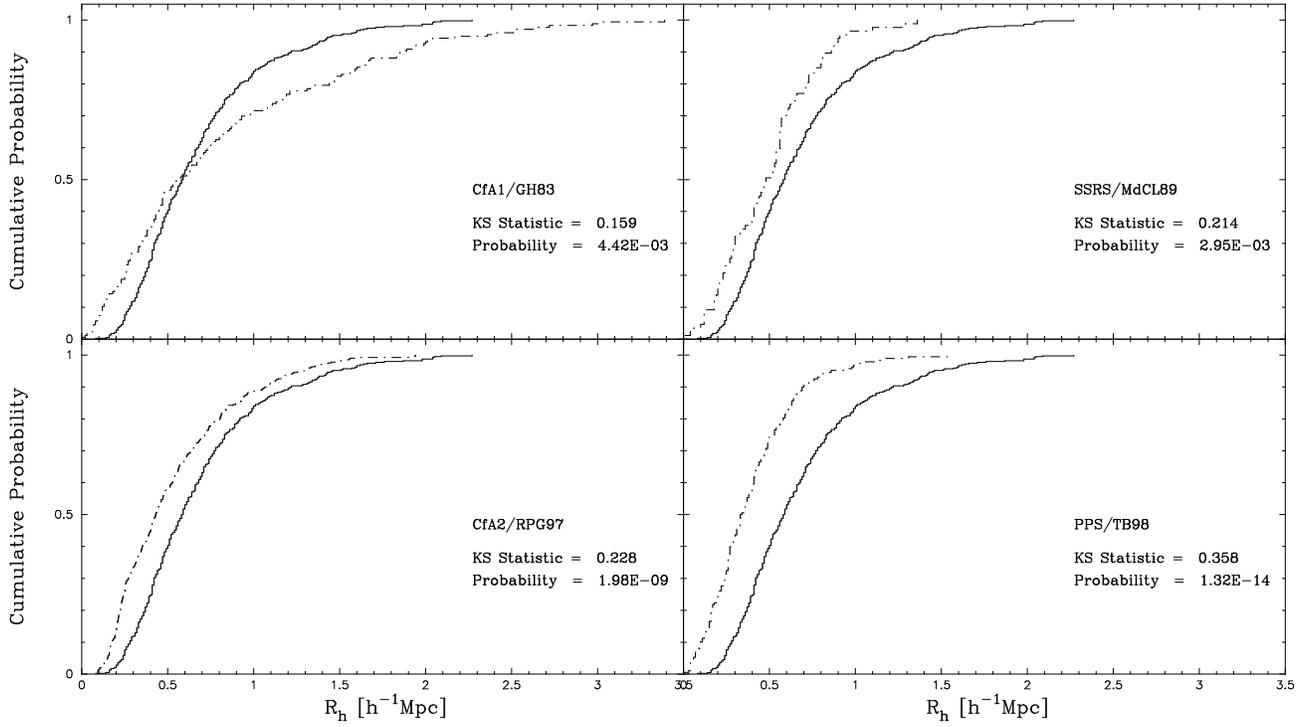}{5.00in}{-90}{75}{75}{-290}{475}
\caption{KS tests comparing the distribution of harmonic radii from
various group catalogues ({\em dot-dashed line\/}) against that from
the full LCRS group catalogue ({\em solid line\/}).}
\end{figure}

\clearpage

\begin{figure}
\plotfiddle{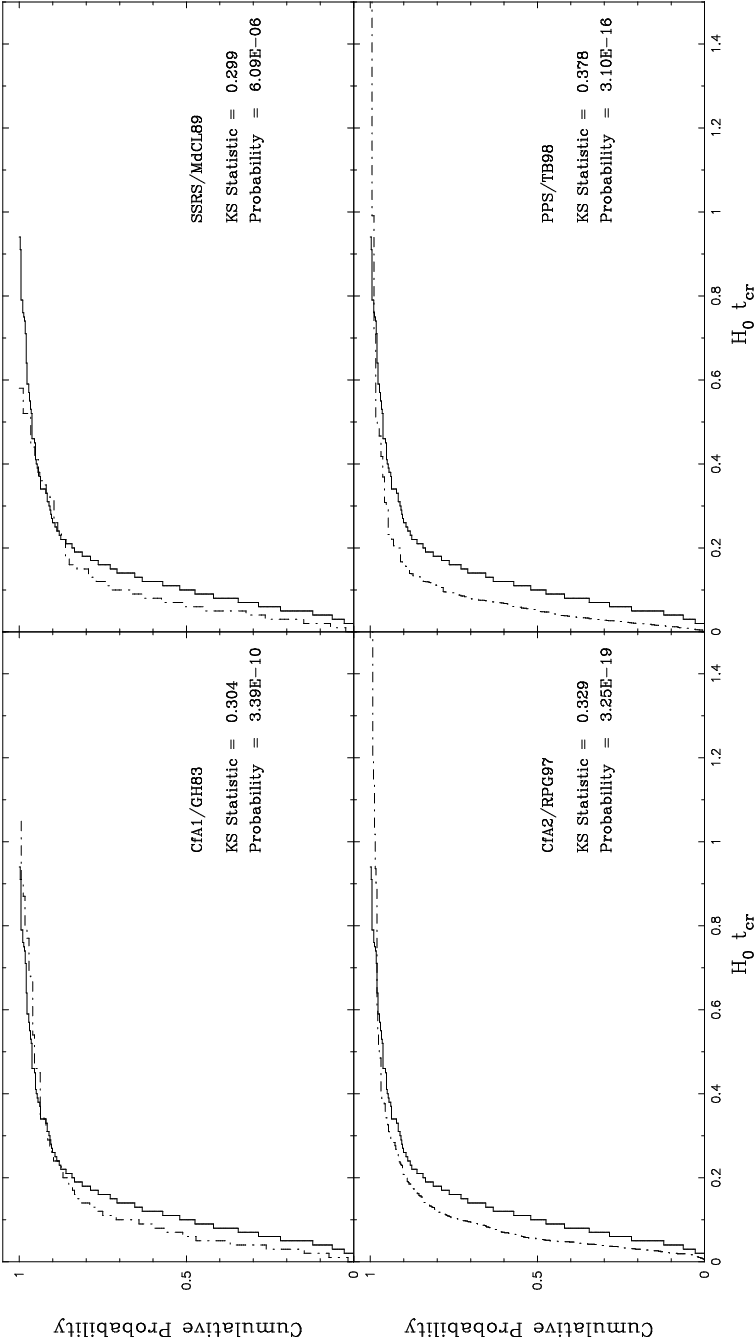}{5.00in}{-90}{75}{75}{-290}{475}
\caption{KS tests comparing the distribution of crossing times from
various group catalogues ({\em dot-dashed line\/}) against that from
the full LCRS group catalogue ({\em solid line\/}).}
\end{figure}

\clearpage

\begin{figure}
\plotfiddle{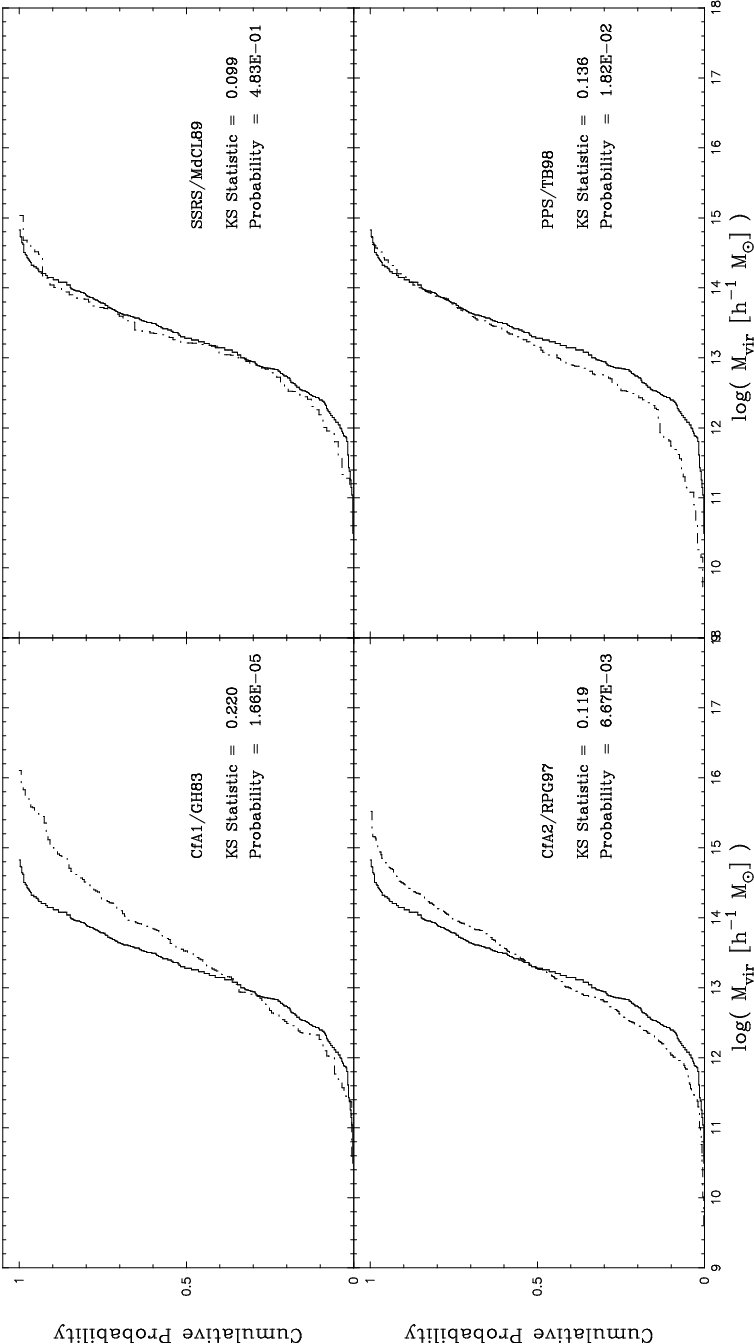}{5.00in}{-90}{75}{75}{-290}{475}
\caption{KS tests comparing the distribution of virial masses from
various group catalogues ({\em dot-dashed line\/}) against that from
the full LCRS group catalogue ({\em solid line\/}).}
\end{figure}

\clearpage

\begin{figure}
\plotfiddle{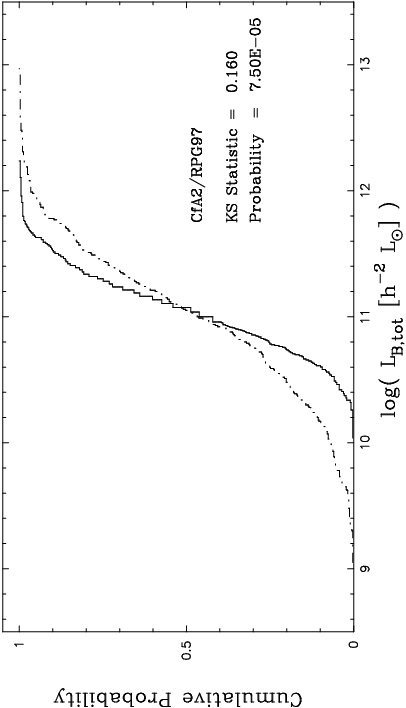}{5.00in}{-90}{75}{75}{-290}{475}
\caption{A KS tests comparing the distribution of de~Vaucouleurs
$B(0)$-band total group luminosities from RPG97 ({\em dot-dashed
line\/}) against that from the full LCRS group catalogue ({\em solid
line\/}).  (The LCRS group luminosities have been converted from LCRS
$R$-band to de~Vaucouleurs $B(0)$-band via equation~39.)}
\end{figure}

\clearpage

\begin{figure}
\plotfiddle{f27.ps}{5.00in}{-90}{75}{75}{-290}{475}
\caption{A KS tests comparing the distribution of de~Vaucouleurs
$B(0)$-band group mass-to-light ratios from RPG97 ({\em dot-dashed
line\/}) against that from the full LCRS group catalogue ({\em solid
line\/}).  (The LCRS group luminosities have been converted from LCRS
$R$-band to de~Vaucouleurs $B(0)$-band via equation~39.)}
\end{figure}

\clearpage

\begin{figure}
\plotfiddle{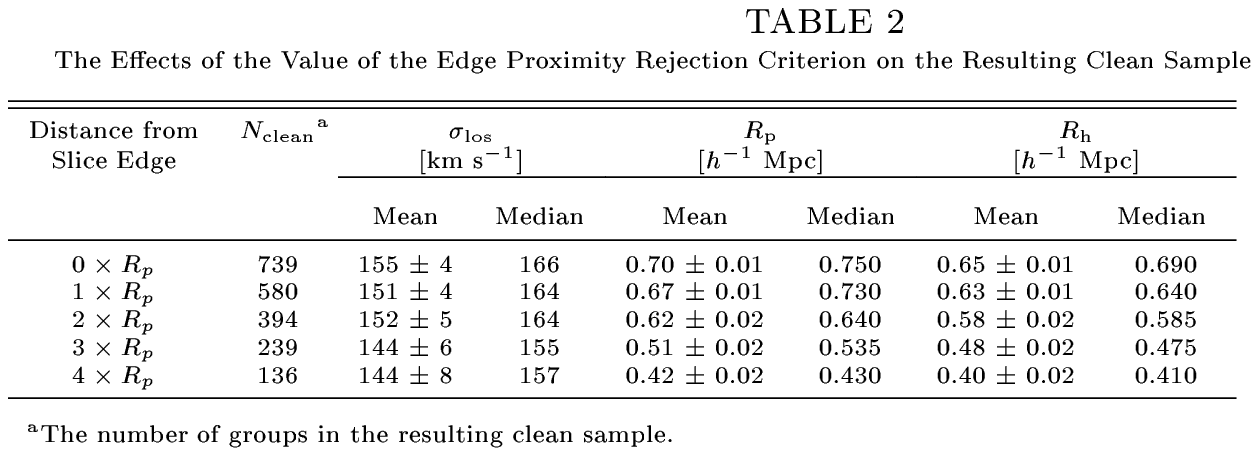}{5.00in}{0}{100}{100}{-300}{-200}
\end{figure}

\clearpage

\begin{figure}
\plotfiddle{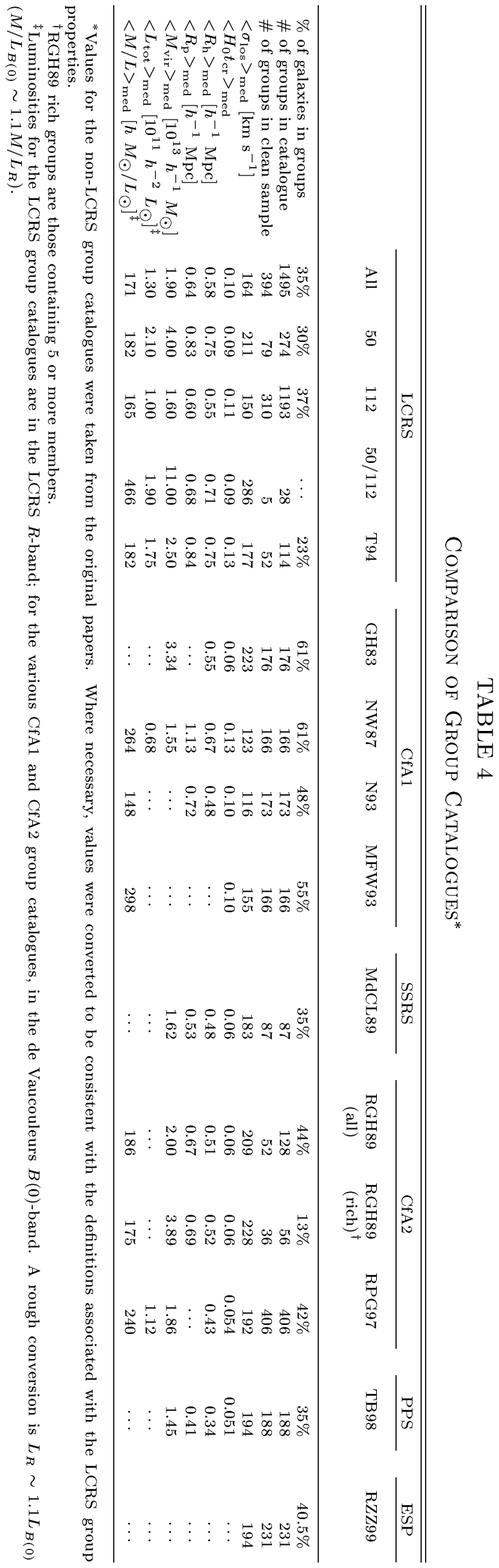}{5.00in}{180}{90}{90}{200}{520}
\end{figure}

\clearpage

\begin{figure}
\plotfiddle{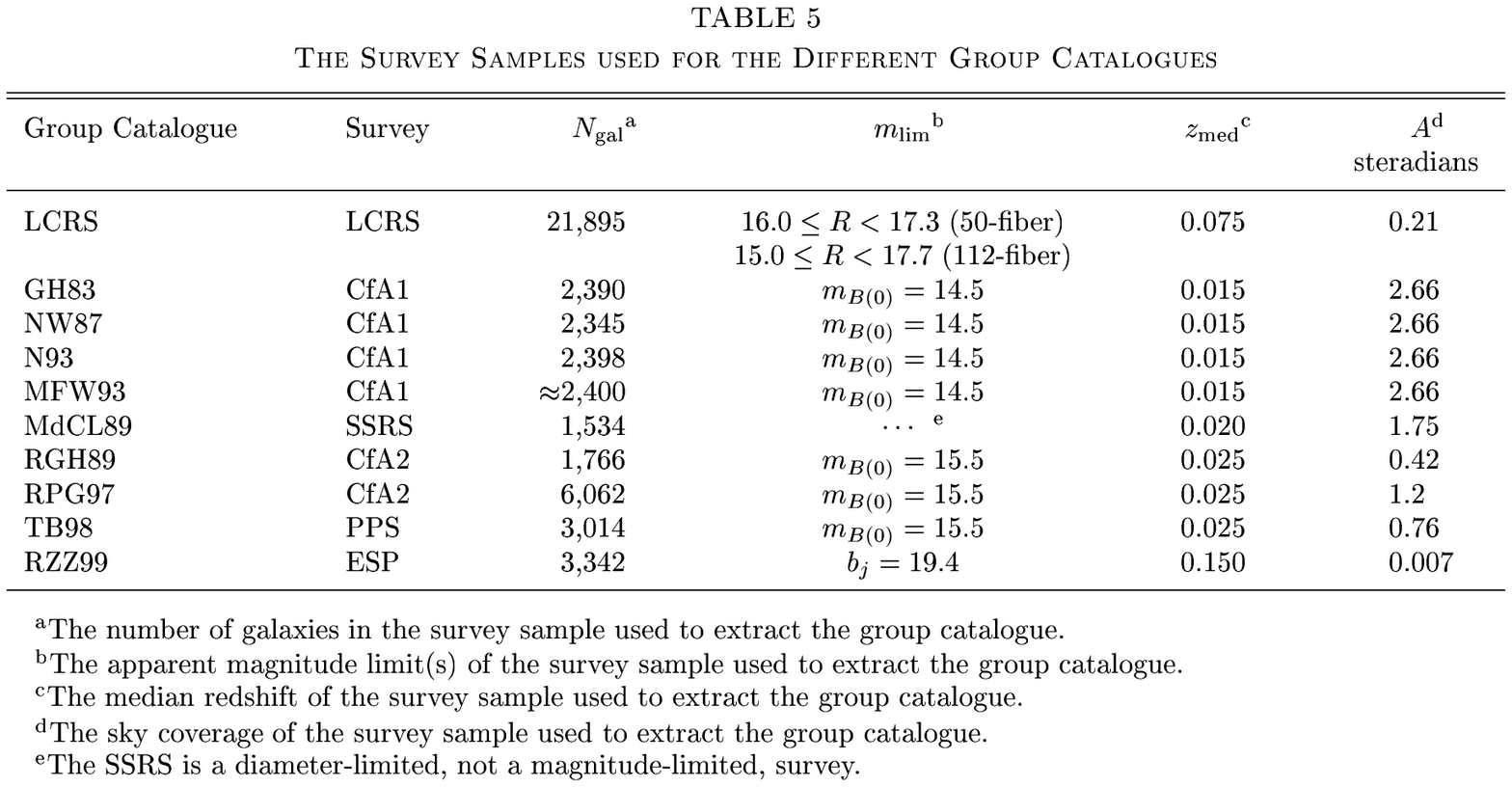}{5.00in}{0}{100}{100}{-300}{-200}
\end{figure}

\clearpage

\begin{figure}
\plotfiddle{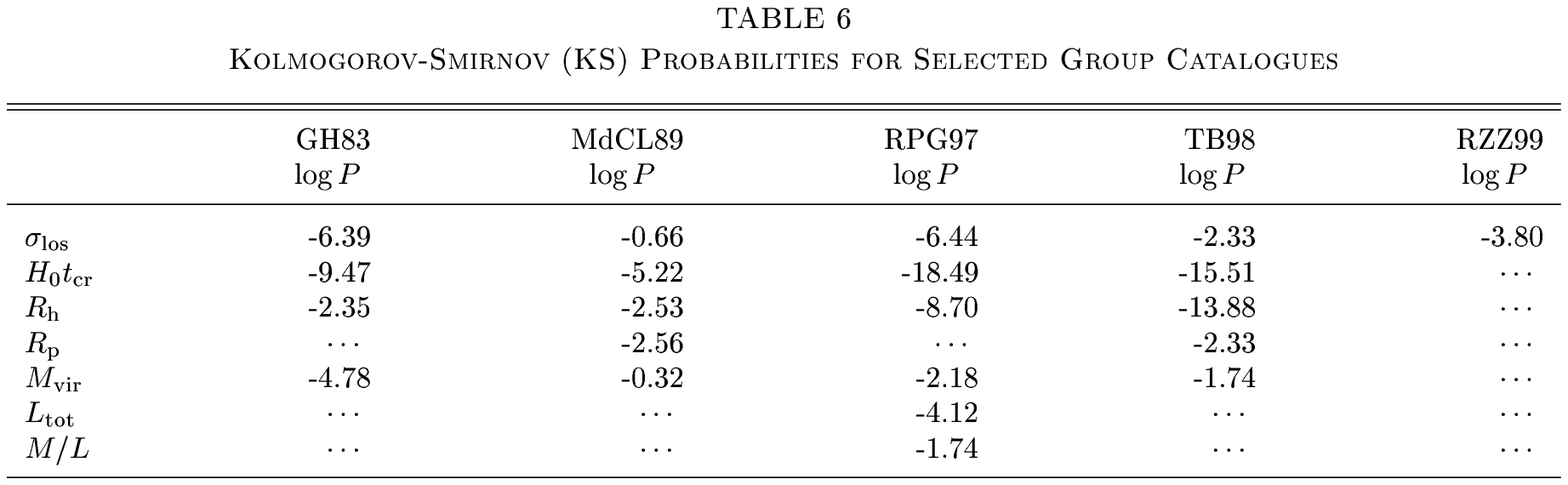}{5.00in}{0}{100}{100}{-300}{-200}
\end{figure}

\clearpage

\begin{figure}
\plotfiddle{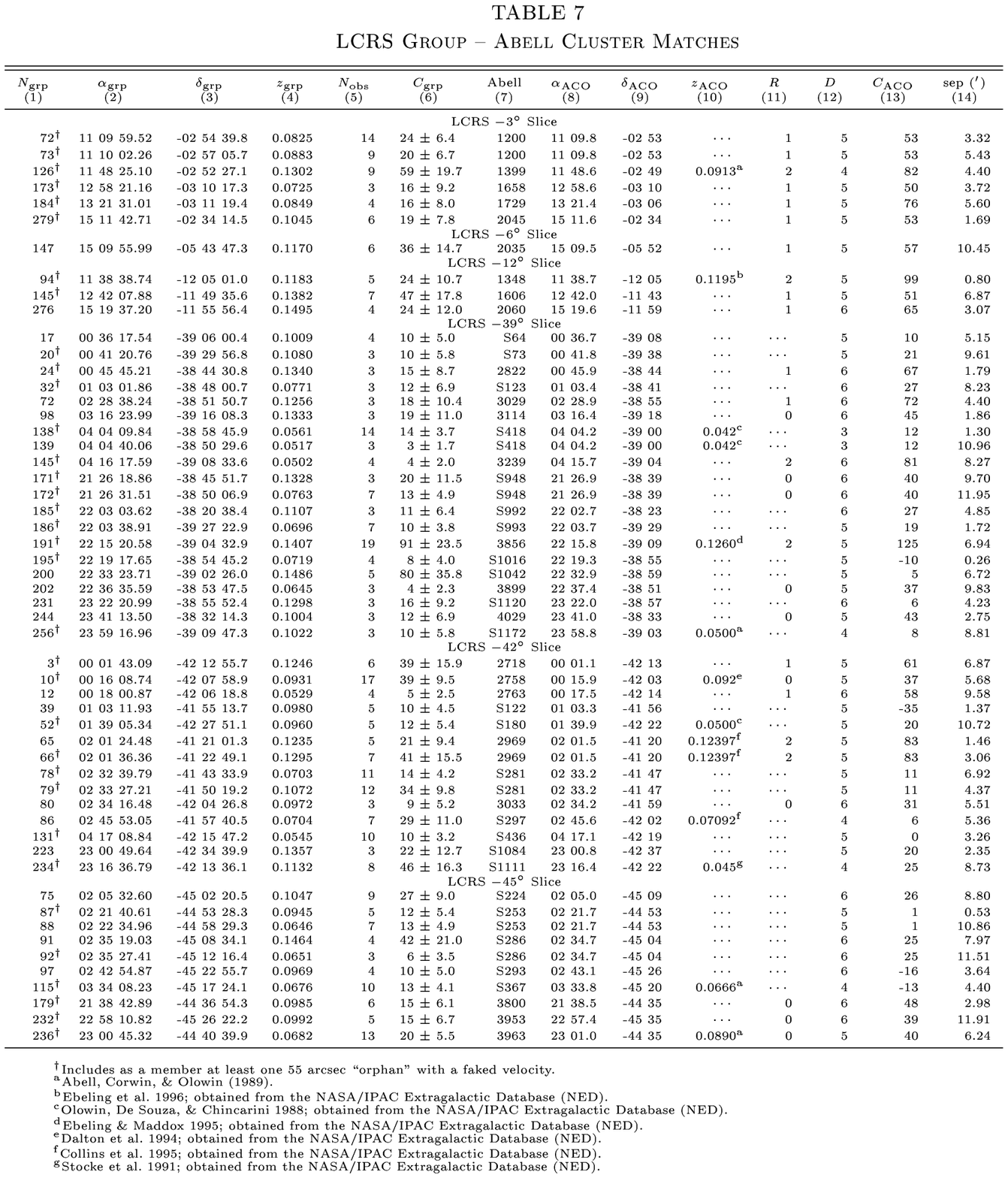}{5.00in}{0}{100}{100}{-300}{-200}
\end{figure}

\end{document}